\documentclass{aa}  
\usepackage{hyperref,xcolor}
\usepackage{comment}
\usepackage[ocgcolorlinks]{ocgx2}
\usepackage{graphicx}
\usepackage{txfonts}
\usepackage{lipsum}
\usepackage{subcaption}                                        
\usepackage{lscape}             
\usepackage{natbib}
\usepackage{placeins}           %

\begin{document}

\hypersetup{allcolors=blue,allbordercolors=blue}
   \title{First application of the optimised next neighbour cleaning to data from imaging atmospheric Cherenkov telescopes}

   \author{M. Kherlakian \inst{1, 2, 3}
        \and G. Maier\inst{1}
        }

   \institute{Deutsches Elektronen-Synchrotron DESY,
Platanenallee 6, 15738 Zeuthen, Germany
            \and Humboldt-Universität zu Berlin, Faculty of Mathematics and Natural Sciences, Newtonstraße 15, 12489 Berlin, Germany \and  Ruhr-Universität Bochum
Astronomisches Institut, Universitätsstr. 150, D-44801 Bochum\\ \email{kherlakian@astro.ruhr-uni-bochum.de} }

  \abstract
   {Imaging atmospheric Cherenkov telescopes (IACTs) cameras are constantly subject to signals from the night sky background and electronic noise during event recordings. Image cleaning methods are typically employed to eliminate pixels that are mostly noise-contaminated and lack sufficient Cherenkov light. In this paper, we present the implementation of the optimised next neighbour (ONN) technique to the Very Energetic Radiation Imaging Telescope Array System (VERITAS) reconstruction pipeline. With this method, cleaning thresholds are derived in a two-parameter space, composed of the minimum total integrated charge of a cluster of pixels, $q_{\rm min}$, and by the maximum difference in the pulse arrival time within the cluster, $\delta_t$. The cleaning contours in $q_{\rm min} - \delta_t$ are adjusted to set a high probability of noise suppression while optimising the retention of pixels containing Cherenkov light. In this paper, we investigate the impact of different levels of image noise contamination on the reconstruction efficiency. A maximum probability of noise cluster retention of $f_{\mathrm{acc}} = 5 \times 10^{-4}$ was found to provide the optimal balance between noise suppression, retention of Cherenkov signals, and preservation of image morphologies. In comparison to the VERITAS double-pass image cleaning method, effective areas are increased at the lower energy range ($\sim$ 60 - 90 GeV) by a factor of up to $\sim$ 25. Energy thresholds are reduced by up to $\sim$ 22\%. Testing the method on Crab Nebula data resulted in an increased number of reconstructed excess events at energies < 1 TeV by a factor of up to $\sim$ 82. Finally, an increase in the differential sensitivity of VERITAS by a factor of up to $\sim$ 8 was observed at $\sim$ 20 GeV. With the addition of BDT-based $\gamma$/hadron separation cuts, the method improves the detection significance of a 20 hour Crab Nebula dataset, while excess event rates are enhanced by $\sim$ 16\%. Comparisons between Monte Carlo simulations and the data do not indicate any additional systematic effects. In summary, the ONN image cleaning provides a reduced energy threshold and a higher rate of event reconstruction. The technique is especially advantageous for the analysis of gamma-ray sources presenting transient and soft spectra.}


   \maketitle
   \nolinenumbers

\section{Introduction}
\label{sec:intro}
Imaging atmospheric Cherenkov Telescopes (IACTs), such as VERITAS~\citep{Holder_2006}, MAGIC~\citep{Garczarczyk2011zz}, and H.E.S.S.~\citep{Hinton_2004} have discovered hundreds of new astrophysical sources in the very-high-energy (VHE, $>$~100 GeV) sky. These instruments are equipped with $\sim 10^2$ m$^2$ reflective mirrors and efficient photodetector cameras~\citep{vtspmt, Fink:2015fia, zorn2019sensitivityimprovementsveryhighenergygammaray} designed to capture the nanosecond Cherenkov pulses induced by gamma- and cosmic-ray extensive air showers (EAS). Advanced analysis techniques have led to an improved cosmic-ray background rejection and enhanced angular and energy resolution~\citep{2013APh4319H, Weekes:2005da}, which significantly boosted the sensitivity of IACTs. With the software and hardware upgrades planned for the Cherenkov Telescope Array Observatory (CTAO), which is next-generation of ground-based gamma-ray observatories at the VHE regime, an increase in sensitivity by a factor of at least 5 is anticipated across the energy range of $\sim$ 20 GeV - 300 TeV~\citep{sensi, abe2023crab}.

During event readout, IACT cameras record not only Cherenkov shower signals from EAS, but also substantial contamination from the night-sky background (NSB; contributions include starlight, zodiacal light, moonlight, airglow effects, and light pollution from urban areas; \citealt{nsb, Preu_2002}). Moreover, cameras also readout electronic noise, such as afterpulsing signals from photomultiplier tubes (PMTs) and spontaneous fluctuations of currents, voltage, and readout or digitisation devices. Overall, NSB rates can reach levels up to 1000 MHz per camera pixel during nights under intense moonlight. If pixels that are dominated by noise photons are not properly removed during event reconstruction, they can introduce significant biases during the image parametrisation and degrade the overall reconstruction performance. To identify and remove such pixels, IACT analysis pipelines apply image cleaning methods designed to retain mostly pixels that contain a significant level of the Cherenkov light from particle showers and to optimise the background rejection. In this paper, we present the application of the optimised next neighbour algorithm (ONN; \citealt{onn}) to the VERITAS array. In brief, ONN is a time-based image cleaning method which efficiently suppresses clusters of pixels induced by the background noise, while allowing lower signal thresholds to be applied.

VERITAS is an array of four 12 m Davies–Cotton IACTs located at the Fred Lawrence Whipple Observatory (FLWO) in southern Arizona (31 40N, 110 57W,  1.3km a.s.l.) that has been in operation since 2007. The array is primarily designed to detect gamma rays in the energy range from $\sim$ 50 GeV up to $>$ 80 TeV. Two major upgrades have contributed to an improved instrument sensitivity; namely, the relocation of one of the telescopes to optimise the array layout~\citep{perkins2009veritas} and the replacement of the 499 camera photomultiplier tubes (PMTs) with high quantum efficiency models \citep{kieda2013gamma, vtspmt}. Major advances have also been made in the analysis pipeline, including the introduction of $\gamma$/hadron separation cuts based on multivariate analysis techniques~\citep{krauseImprovedGammaHadron2017} and the yearly throughput calibration~\citep{sfactorvts}.

This paper is organised as follows: In Sect.~\ref{sec:veritas}, we present an overview of the VERITAS double-pass cleaning algorithm and the ONN image cleaning procedure, with a description of the method workflow. Its performance is evaluated using simulations and observational data in Sects.~\ref{sec:sim_performance} and~\ref{sec:data_performance}, respectively. The impact on the differential sensitivity is discussed in Sect.~\ref{sec:sensitivity}, while our conclusions are presented in Sect.~\ref{sec:conclusions}.

\section{Image cleaning for VERITAS}
~\label{sec:veritas}

\noindent The core principle behind time-based image cleaning methods relies on exploring the differences in temporal and spatial behaviours of noise pulses and Cherenkov signals. Cherenkov pulses fall within the $\sim$ 2-10 ns range \citep{Aharonian_2004}, depending on the properties of the primary particle and on the design of the IACT camera. The Cherenkov light pool strikes adjacent pixels within shorter time intervals (from within the time resolution of the instrument; in the case of VERITAS: $10^{-3}$ ns up to a few). For this reason, images from EAS exhibit strong temporal correlation and are spatially clustered. On the other hand, NSB and electronic noise fluctuations follow Poisson statistics and exhibit no spatial or temporal correlation. The spatial correlation (i.e, neighbouring) of pixels with signals induced by noise decreases rapidly with increasing pixel charge. This allows pixel timing constraints to improve the discrimination between shower signals and noise.
 
VERITAS employs a three-level trigger system to reduce the recording rate of noise~\citep{trigger}. At the first level (\textit{L1}), a constant fraction discriminator (CFD) is triggered when a pixel signal exceeds 50 mV. The telescope level trigger (\textit{L2}), requires at least three adjacent pixels within a single telescope to register \textit{L1} triggers in a 6 ns time coincidence window. The last and third level (\textit{L3}), known as the array trigger, requires two telescopes (or more) to register an \textit{L2} trigger within a 50 ns window. To estimate the baseline noise rate, VERITAS records pedestal events at a rate of 1 Hz during the data acquisition process~\citep{ped}. These pedestal events readout the pixel charges within the time range of 32 ns in the absence of an \textit{L3} trigger. This allows the monitoring of pixel charges under purely noise conditions. To account for temporal fluctuations in the noise, such as those caused by the rising or setting of the Moon, the pedestal variances are averaged over three-minute intervals.

\subsection{The VERITAS double-pass method}

Conventionally, VERITAS employs the double-pass cleaning method~\citep{doublepass, Holder_2006}, which involves the application of high signal thresholds for the pixel brightness (i.e. the total integrated charge in a pixel within a time range of 32 ns) to effectively eliminate pixels mostly contaminated by noise~\citep{Lessard_2001, Holder_2006}. These high thresholds are very efficient in removing upward noise fluctuations and are applied in the following manner: an initial filtering selects pixels within 5$\sigma$ above the standard deviation of the pedestal (known as \textit{pedvar}). These pixels are known as the core or picture image pixels. Following, a 2.5$\sigma$ threshold is applied for neighbours of the core pixels. Only picture pixels with at least one neighbour passing the 2.5$\sigma$ criterion are included in the final image. This topological condition guarantees that spurious islands of upward noise fluctuations caused by bright stars, background emission, or secondary fainter Cherenkov emission are removed. After the initial cleaning, the time gradient, $t_\mathrm{grad}$, of the pulse arrival time is estimated along the major axis of the image and used to relocate a shorter summation window of 24 ns for the charge integration. This is followed by a second pass of the 5$\sigma$/2.5$\sigma$ brightness thresholds. Improvements in this technique have been achieved by eliminating outlier camera pixels containing signals from afterpulsing effects. These outliers are responsible for distorting the Hillas parametrisation of the images~\citep{Hillas1985, denauroisAnalysisMethodsAtmospheric2006}, and removing them has significantly improved the sensitivity of the telescopes.

\subsection{ONN cleaning}
\label{sec:onn}

Although the traditional image cleaning techniques based on charge thresholds, together with its several variations $-$ for example, as implemented in \textit{ctapipe}~\citep{ctapipe-icrc-2023} and \textit{cta-lstchain}~\citep{lstchain_adass_2020} and NSB adaptative cuts~\citep{abe2023crab, Sajjad} $-$ have been extensively validated by the third-generation of IACTs, one of its main disadvantages is the suppression of showers at low ($\sim$ GeV) energies. The images produced by these showers are typically of low pixel brightness and fail to meet the charge threshold criteria. This limitation motivates the use of alternative image cleaning methods that exploit additional information, such as the temporal coherence of Cherenkov shower signals, to improve the reconstruction efficiency at low energies. The incorporation of time-based cleaning algorithms, such as MAGIC~\citep{shaydukNewImageCleaning2005}, H.E.S.S.~\citep{hessic}, and LST~\citep{abe2023crab}, have yielded enhanced sensitivity. Simulation studies for CTAO have shown that the ONN method offers an enhanced event retention at low energies~\citep{onn}. In this work, we show that it provides an improved array sensitivity at low energies in comparison to the VERITAS standard double-pass~\citep{doublepass, Holder_2006} method, which makes it ideal for detecting faint, soft and transient astrophysical signals. This technique is also expected to result in larger effectiveness of reconstruction in comparison to methods with fixed intervals for pulse arrival time coincidences, a detailed in the literature ~\citep{shaydukNewImageCleaning2005, abe2023crab}.

The standard 5$\sigma$/2.5$\sigma$ thresholds can be lowered by requiring short arrival-time differences between pulses in neighbouring pixels. The ONN method exploits this behaviour by applying dynamic selection cuts in the parameter space defined by the minimum pixel charge, $q_\mathrm{min}$, and the maximum arrival-time difference, $\delta t_\mathrm{max}$, within a cluster of $n_{NN}$ neighbouring pixels. In this work, clusters with multiplicities $n_{NN}$ = 2, 3 and 4 pixels are considered. The cuts are chosen such that the expected rate of surviving noise pixels in the image remains constant. The rate of surviving noise pixels is described by the following equation,

\begin{center} \begin{equation} \centering \mathcal{R}_{\text{acc}}(n_\mathrm{pix}, q_\mathrm{min}, \delta t_\mathrm{max}) = C_{n_{NN}} \, \left( \mathcal{I}(q_\mathrm{min}) \right)^{n_{NN}} \, \delta t_\mathrm{max}^{n_{NN}-1}, \label{eq:racc} \end{equation} \end{center}

where

\begin{itemize}

    \item $n_{NN}$ denotes the number of pixels in the cluster;

    \item $q_\mathrm{min}$ is the minimum pixel brightness within the cluster of $n_{NN}$ pixels,
 \begin{equation*}q_\mathrm{min} = min\left( ~q_1, q_2, ..., q_{n_{NN}} ~\right);\end{equation*}

    \item $\delta_{t_\mathrm{max}}$ is the maximum pulse arrival-time difference within the cluster,
    \begin{equation*}
        \delta_{t_\mathrm{max}} = max\left( | t_\mathrm{i} - t_\mathrm{j} | \right), \textrm{with $i, j \in$ $1, 2, ..., n_{NN}$; $i$ $\neq$ $j$};
    \end{equation*}
    \item $C_{n_{NN}}$ is a constant factor that accounts for all possible geometrical arrangements of a cluster containing $n_{NN}$ neighbouring pixels in each VERITAS camera, composed of 499 pixels with hexagonal shape;
    
    \item $\mathcal{I}(q_\mathrm{min})$, known as individual pixel rate (IPR), describes the rate at which noise fluctuations induces pixel brightness up to $q_\mathrm{min}$;
    
    \item  $\mathcal{R}_{\text{acc}}(n_{NN}, q_\mathrm{min}, \delta t_\mathrm{max})$, known as the accidental cluster rate corresponds to the rate of noise-induced pixels surviving the cleaning procedure. This parameter is fixed and can be set to values as low as desired. Lower values of $\mathcal{R}{\mathrm{acc}}$ result in more stringent cleaning cuts and stronger suppression of noise. In practice, instead of dealing directly with rate ranges, we defined a maximum probability of noise cluster retention, $f_{\mathrm{acc}}$, with the accidental cluster rate being defined as the ratio of $\mathcal{R}_{\mathrm{acc}}$ by the length of the signal integration window.
\end{itemize}

For a constant value of the accidental cluster rate, the dynamical contours in the $q_\mathrm{min}$ - $\delta t_\mathrm{max}$ can be drawn as

\begin{multline}
    \delta t_\mathrm{max}(q_\mathrm{min}, n_{NN}, \mathcal{R}_{\text{acc}}) = \\ \vspace{0.1cm}
    \exp\left[ \frac{1}{n_{NN} - 1} \ln\left( \frac{\mathcal{R}_{\text{acc}}}{C_{n_{NN}} \mathcal{I}^{n_{NN}}(q_\mathrm{min})} \right) \right].
    \label{eq:dt-q}
\end{multline}

\begin{figure}
   \centering
   \includegraphics[width=0.77\hsize]{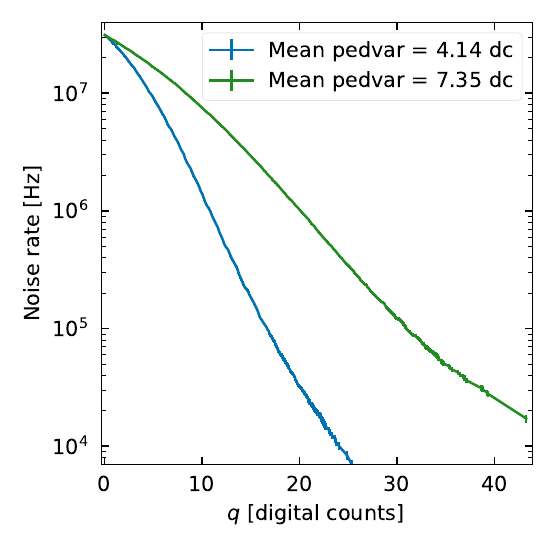}
      \caption{Noise rate as a function of pixel brightness. The rate is given as the average over all cameras pixels for VERITAS telescope T1. The curve in green shows the noise rate for an average pedestal variance of 7.35 dc, while the blue curve, for 4.14 dc. One digital count is equivalent to approximately 5 photoelectrons.}
         \label{fig:IPR}
   \end{figure}

We refer to Appendix \ref{app:contours} for a discussion on Eq. \ref{eq:racc} for different cluster multiplicities. The noise rates (IPRs) for a given observation are constructed by averaging the pedestal measurements within the VERITAS readout window within all pixels in each telescope camera. Observations taken under higher NSB conditions will display higher IPR levels, as seen in Fig. \ref{fig:IPR}, which presents a typical IPR at two different noise baseline levels. 

\subsubsection{Dynamical threshold cuts in the $q_\mathrm{min}$ - $\delta t_\mathrm{max}$ parameter space}

\begin{figure}[h!]
   \centering
   \includegraphics[width=0.77\hsize]{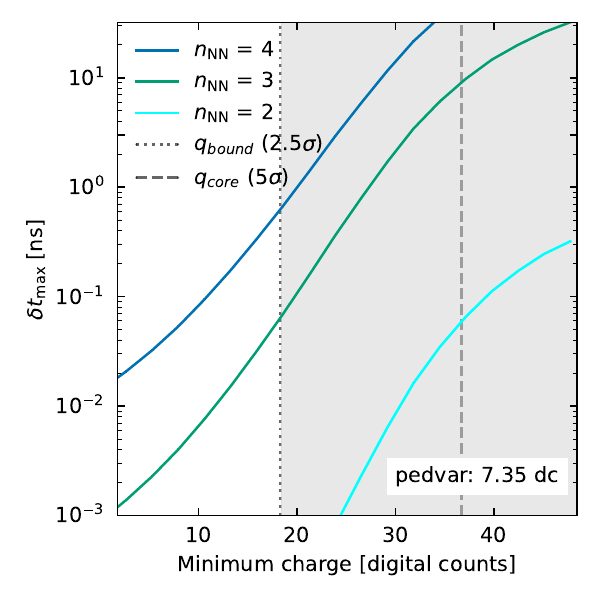}
      \caption{Dynamical threshold cuts given in the $q_\mathrm{min}$ - $\delta t_\mathrm{max}$ parameter space for the VERITAS telescope T1. The contours are shown for the following cluster multiplicities: $n_{NN}$ = 2 (cyan), 3 (green), and 4 (violet). The maximum allowed $\delta t_\mathrm{max}$ spans from 0 up to the size of the readout window (32 ns). The double-pass cleaning thresholds are shown as the grey lines for core (dashed) and boundary pixels (dotted). The gray shaded region represents the allowed parameter space for the nominal method, which is independent from the group multiplicity.}
         \label{fig:contours}
\end{figure}

Figure \ref{fig:contours} shows the dynamical thresholds established via Eq. \ref{eq:dt-q} for $f_{\mathrm{acc}}$ = 5 $\cdot$ 10$^{-4}$ (see Sect. \ref{sec:racc} for a discussion on the choice of $f_{\mathrm{acc}}$) and cluster multiplicities $n_{NN}$ = 2, 3, and 4 for an observation taken under an average pedvar of 7.35 dc. Clusters of pixels with properties below the threshold curves have a probability lower than $f_{\mathrm{acc}}$ of being noise induced. On the other hand, clusters characterised by the properties given above have a probability higher than $f_{\mathrm{acc}}$ and are discarded during the cleaning. Since IPRs are measured via pedestal events, they adapt to the noise conditions of each observations. At periods of intense moonlight, more stringent contours are derived. For comparison, we show the nominal 5$\sigma$/2.5$\sigma$ cuts for the same observation. These are given as vertical lines at positions of 18.38 p.e. and 36.75 p.e. for picture and neighbouring pixels, respectively. The parameter space outside the grey shaded region would be overlooked in the nominal technique.

\subsubsection{Algorithm workflow}

With the dynamical thresholds established, the workflow of the ONN image cleaning proceeds as follows:

\begin{enumerate} 

\item Initial filtering: A preliminary brightness threshold of 2 d.c. is applied at the single-pixel level to avoid selecting pixels with null charge;

\item Search for $n_{NN}$ = 2: pairs of neighbouring pixels satisfying the $n_{NN}$ = 2 contour are saved to compose the final image; 

\item Search for $n_{NN} = 3$: following, a search is conducted on pixels adjacent to the pairs that passed the $n_{NN}$ = 2 threshold. If the resulting three-pixel cluster satisfies the $n_{NN} = 3$ threshold, the new pixel is incorporated. We note that the threshold for three-pixel clusters is less strict than for pairs;
\item Search for $n_{NN} = 4$: in a similar manner, the search extends to pixels adjacent to the three-pixel clusters. If a four-pixel configuration meets the $n_{NN} = 4$ contour criterion, it is added to the image.
\end{enumerate}

\begin{figure}
   \centering
   \includegraphics[width=0.97\hsize]{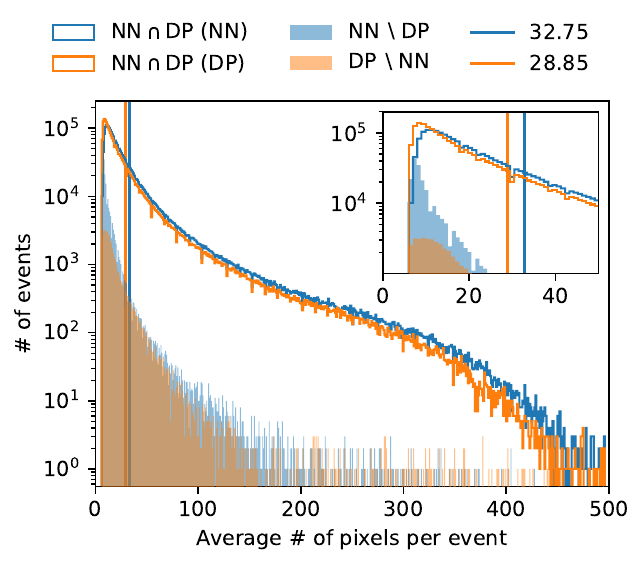}
      \caption{Total number of pixels, $n_{NN}$, for triggered simulated events for the ONN (blue) and double-pass (DP, orange) cleaning methods. Both histograms are constructed by averaging the number of pixels  within all images reconstructed by the four telescopes of the VERITAS array. Step histograms ($\mathrm{NN}\cap\mathrm{DP}$) represent distributions for reconstructed simulated events that pass data quality cuts for both methods. Filled histograms ($\mathrm{NN}\setminus\mathrm{DP}$ or $\mathrm{DP}\setminus\mathrm{NN}$) represent events reconstructed and above quality cuts by only one method. Horizontal lines represent the mean of the step histograms.}
         \label{fig:npixevent}
\end{figure}

The implementation was done within the \textit{EventDisplay} framework~\citep{maierEventdisplayAnalysisReconstruction2017}, an open-source analysis package designed for the reconstruction and display of events collected by VERITAS and the CTAO. When optimising the removal of noise pixels, cleaning algorithms should also guarantee that the maximal number of pixels containing the true shower signal are kept for the reconstruction. However, in practice, it has also been observed that the telescope sensitivity reaches a plateau with respect to the number of true Cherenkov pixels in the image~\citep{onn}. At a certain point, a higher number of reconstructed pixels containing the true signal does not necessarily provide a better performance. This occurs because pixels farther from the image core contain weaker Cherenkov signals and should have a lower effect on image parametrisation.  For this reason, multiplicities above four are not considered, as they do not substantially improve the cleaning process and would significantly increase computational time. Lowering the minimum pixel brightness allowed for image reconstruction, as seen in Fig. \ref{fig:contours}, allows for the reconstruction of faint images originating from low-energy showers that would otherwise be discarded. This is exemplified in Fig. \ref{fig:npixevent}, which shows the distribution of the average number (within all four telescopes) of pixels from simulated gamma-ray images which survive the cleaning. For events reconstructed by both the ONN and baseline methods (NN$\cap$DP), the distribution obtained with ONN is shifted towards larger average number of pixels. This is expected, since the ONN is able to retain additional pixels that are rejected by the standard double-pass cleaning. If we are considering events reconstructed exclusively by ONN (NN$\setminus$DP), we would notice that the ONN reconstructs more images with few pixels ($\sim$ 10 -15) in comparison to events reconstructed exclusively with the baseline method (DP$\setminus$NN).

\subsubsection{The accidental cluster rate}
\label{sec:racc}

The accidental rate, $\mathcal{R}_{\text{acc}}$, can, in principle, be set to arbitrarily low values. However, doing so would increase pixel rejection, including for pixel clusters containing light from the Cherenkov pulses. Not only would this suppress gamma-ray events, but it would also negatively affect the parametrisation of shower images and degrade $\gamma$/hadron separation. Conversely, setting a lenient accidental cluster rate should increase noise contamination and also introduce biases in the images. To find a suitable range for the accidental cluster rate, we performed Monte Carlo simulations of gamma-ray events and clean them assuming different values of $f_{\text{acc}}$.

Showers were simulated with the CORSIKA (COsmic Ray SImulations for KAscade) program~\citep{corsika} and channelled into the CARE\footnote{\href{https://github.com/nepomukotte/CARE}{https://github.com/nepomukotte/CARE}} or GrOptics\footnote{\href{https://github.com/groptics/GrOptics}{https://github.com/groptics/GrOptics}}  packages, which simulate the telescope response. The simulated images pass through the \textit{EventDisplay} pipeline, which performs the steps of calibration, image cleaning, and image parametrisation. 

\begin{figure}
    \centering
     \includegraphics[width=0.77\hsize]{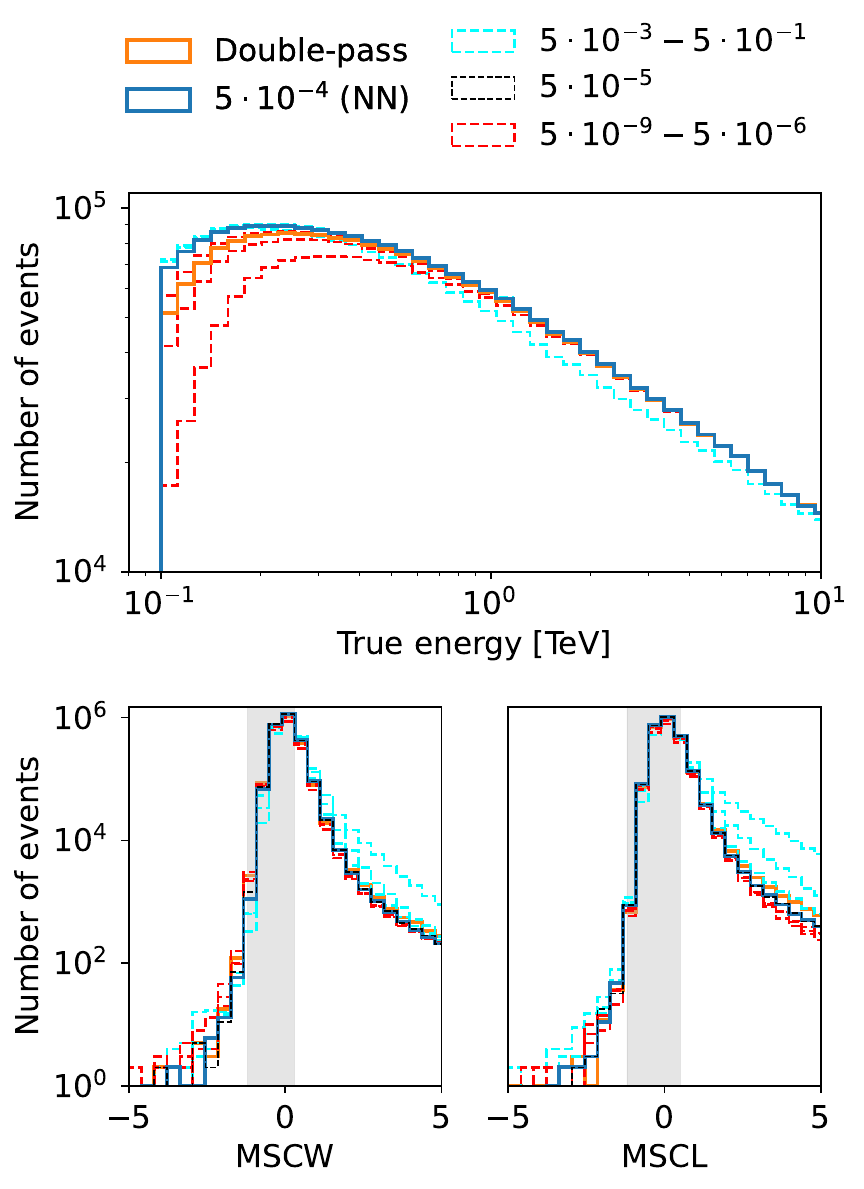}
    \caption{Energy spectrum of simulated gamma-ray events (\textit{top}), distribution of simulated MSCW (\textit{bottom-left}) and distribution of simulated MSCL (\textit{bottom-right}) for different values of $f_{\text{acc}}$. Simulations were performed at $\phi_z$ = 20 degrees and $\eta$ = 200 MHz. In cyan: $5\cdot 10^{-3} < f_{\mathrm{acc}} < 5\cdot 10^{-1}$. In red: $5\cdot 10^{-9} < f_{\mathrm{acc}} < 5\cdot 10^{-6}$. Blue: histograms for the nominal cleaning. Orange: histograms for the ONN image cleaning, with the chosen $f_{\mathrm{acc}} = 5\cdot 10^{-4}$. The also suitable value of $f_{\mathrm{acc}} = 5\cdot 10^{-5}$ is shown by the black dashed line. For reference, the grey shaded areas represent the supersoft box cuts for $\gamma$/hadron separation.}
    \label{fig:fprob}
\end{figure}

Figure \ref{fig:fprob} presents the number of reconstructed simulated gamma-ray showers assuming the following values of $f_{\text{acc}} = 5 \cdot 10^{-i}$ for $i \in [-9, -7, -6, -5, -4, -2, -3, -1]$. The simulations were produced for showers arriving at a zenith angle of 20 degrees and assuming an NSB rate of 200 MHz (same comparisons but for simulations performed at zenith angles of 50 degrees can be found in Fig. \ref{fig:fprob_50}). To understand the effects of different accidental cluster rates on the image shape, we also show the reconstructed mean scaled width (MSCW) and mean scaled length (MSCL) parameters, which are derived via the Hillas method and have the highest impact on $\gamma$/hadron separation. For comparison, we present the same distributions under the nominal double-pass cleaning (plus afterpulsing correction and from now on referred to only as a so-called double-pass). 

For $f_{\mathrm{acc}}$ $>$ 5 $\cdot$ 10$^{-3}$, the image parameters deviate strongly from the nominal distribution, indicating they are highly distorted due to the retention of several noise groups. On the other hand, for $f_{\mathrm{acc}}$ $\lesssim$ 5 $\cdot$ 10$^{-6}$, the cleaning thresholds are too strict, leading to a reduction in the number of reconstructed events in comparison to the nominal method. Probabilities within $5\cdot 10^{-5} \lesssim f_{\mathrm{acc}} \lesssim 5\cdot 10^{-4}$ offer the optimal range between noise suppression and event retention, with no image distortions. For results presented in this paper, we chose the probability of $f_{\mathrm{acc}}$ = 5 $\cdot$ 10$^{-4}$. This choice results in about five noise clusters in every 10000 shower images per cluster multiplicity (i.e. about 0.05\% of images will be contaminated). In a typical data taking session of 30 minutes, for which a trigger rate of $\sim$ 100 Hz is expected, about 270 images (out of $\sim$ 180000) should contain a fake cluster when considering multiplicities of $n_{NN}$ = 2, 3, and 4 pixels.

\subsubsection{Impact of the accidental cluster rate on pedestal events}

To test the effectiveness of the noise cluster rejection, $\sim$ 22500 pedestal events (resulting in $\sim$ 90000 images considering all four telescopes of the array) were cleaned with the ONN method. These events are taken in the absence of an array trigger and offer a good representation of noise conditions. After pedestals are cleaned, we analysed the survival rate of clusters with three multiplicities $n_{NN}$ = 2, 3, and 4. A total of 104 clusters are found (within all four telescopes), resulting in an effective survival probability of $f_{\mathrm{acc}}$ =  104/(3 $\cdot$ 90000) $\simeq$ 4$\cdot$10$^{-4}$, which is expected. The nominal cleaning results in a similar survival rate of pixels within pedestal images of $\sim 4.5 \cdot 10^{-4}$, which further supports the choice for the accidental rate. A total of 65 images contain noise clusters due to the $n_{NN}$ = 4 contour (i.e. the most lenient one). Cleaning $\sim$ 16000 pedestals with a set probability of $f_{\mathrm{acc}}$ = 5$\cdot$10$^{-3}$ or 0.005\% ($f_{\mathrm{acc}}$ = 5$\cdot$10$^{-1}$ or 50\%)  results in $\sim$ 210 ($\sim$ 9300) images containing a fake group. This translates into an effective noise survival rate of $\simeq$ 4.4$\cdot$10$^{-3}$ (0.3), which also confirms the theoretically expected survival rate of noise. A histogram of the total number of surviving pixels is shown in Fig. \ref{fig:ntubes_peds}.

\section{Performance study for simulated gamma rays}
\label{sec:sim_performance}
The performance of the ONN image cleaning method was first evaluated with Monte Carlo simulations of gamma-ray showers. These were produced with CORSIKA according to a differential energy spectrum of $dN/dE \propto E^{-2}$. To compare the event reconstruction power of the ONN technique with the nominal one, we calculate the total effective detection area of the array. Total effective detection areas, $A_{\mathrm{eff}}\left( E, \theta_{w},  \eta, \phi_{z}, \phi_{a} \right)$ = $\pi R^{2}_{\mathrm{eff}} \left( N^{\textrm{eff}}_{\gamma} / N^{\textrm{total}}_{\gamma} \right)$, are proportional to the ratio of gamma rays that pass $\gamma$/hadron separation cuts, $N^{\textrm{eff}}_{\gamma}$, to the total number of simulated showers, $N^{\textrm{total}}_{\gamma}$. This ratio is then multiplied by a circular area, $\pi R^{2}_{\mathrm{eff}}$, perpendicular to the shower axis and of a size of $\sim$ 1.5 $\cdot$ $10^6$ m$^2$ (with a radius of up to $R_{\mathrm{eff}}$ = 700 meters), at which the impact parameter of the shower is simulated. Effective areas are energy-dependent and produced in discrete mode in terms of the wobble offset, $\theta_{w}$, NSB levels, $\eta$, zenith, $\phi_{z}$, and azimuth, $\phi_{a}$. The EventDisplay package provides the rate of gamma-ray events, $N^{\textrm{eff}}_{\gamma}\left( E, \theta_{w},  \eta, \phi_{z}, \phi_{a} \right)$, that pass the separation cuts.

The reconstructed energy of gamma-ray showers, $E_{\rm rec}$, is estimated via the method of lookup tables~\citep{sfactorvts}, which are computed through simulations and provide the relation between shower energy, distance of the shower impact position to each telescope and the image brightness (i.e. the summation of all pixel brightness that survive cleaning). Lookup tables are constructed as a function of NSB level and zenith and azimuth angles. The arrival direction of showers is estimated via stereoscopic techniques~\citep{Hofmann_1999}. It corresponds to the mean intersection point between pairs of telescope images. Each pair is weighted by the image brightness, elongation and relative angle. Images that have approximately parallel orientation are down-weighted~\citep{maierEventdisplayAnalysisReconstruction2017}.

Since effective areas are highly dependent on the parameters listed above, we decide to compare the performance for configurations involving different zenith angles and noise levels, which are the most critical ones. For instance, effective areas at low energies are generally lower at small elevation angles because the showers need to transverse a higher atmospheric length, leading to a loss of showers that reach a maximum photon density above the altitude of the array.  We only give the results for the standard 0.5$^{\circ}$ wobble mode. Because we were solely interested in analysing the differences in performance introduced by the image cleaning methods, we applied standard box $\gamma$/hadron separation cuts optimised for supersoft sources based on Crab Nebula data cleaned with the nominal method (noting that the addition of cuts optimised with multivariate analysis techniques \citep{krauseImprovedGammaHadron2017} are analysed in Sect. \ref{sec:bdt}). The chosen $\gamma$/hadron separation cuts are optimised for super-soft sources (i.e. assuming a power-law spectral index of -3) and should be the default when analysing the cleaning effects at low energies. The same $\gamma$/hadron cuts can then be applied to events cleaned with the ONN and baseline methods.

Figure \ref{fig:effarea} presents the effective areas at the  following configurations:
\begin{itemize} 
\item $\phi_{z} \in$ [20$^{\circ}$, 40$^{\circ}$, 60$^{\circ}$];
\item $\eta \in$ [75 MHz, 200 MHz, 400 MHz].
\end{itemize}

\begin{figure}
    \centering
    \includegraphics[width=0.99\hsize]{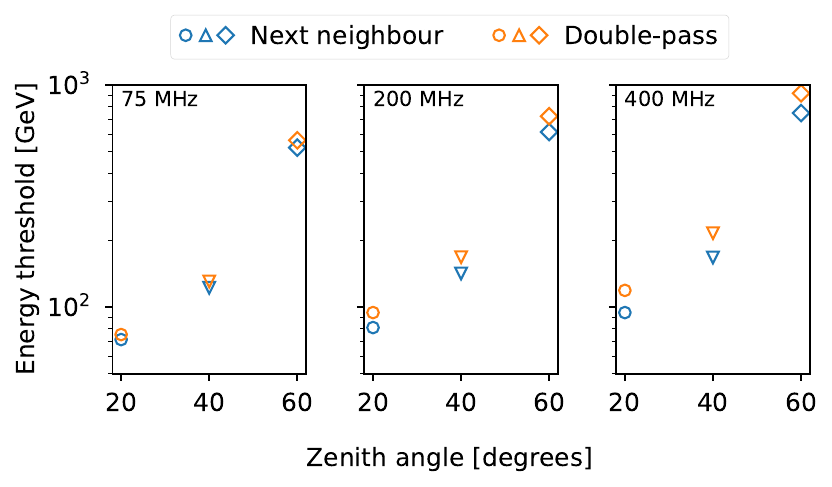}
    \caption{Energy threshold as a function of zenith angles and noise levels for the ONN (blue markers) and baseline cleaning (orange markers).}
    \label{fig:energy_threshold}
\end{figure}

We note that the ONN image cleaning method generally results in higher effective areas, particularly at low energies. This becomes evident from the ratio plots present in the lower panels of Fig. \ref{fig:effarea}, where we see an increase in effective areas by a factor of up to $\sim$ 18.1 (11.7, 25.6) in the first common energy bin for $\phi_z$ = 60$^\circ$ (20$^\circ$, 40$^\circ$). It also becomes evident that for most configurations, effective areas extend to energies that remain inaccessible by the nominal method. Both improvements are primarily due to the higher retention level of low-energy gamma-ray showers, $N^{\textrm{eff}}_{\gamma}\left( E \lesssim \; 100 \mathrm{GeV}\right)$, with the ONN. In addition, the energy threshold above which spectral analyses are performed, defined as the energy at which the effective area reaches 10\% of its maximum, is also generally lower with the ONN method, as shown in Fig. \ref{fig:energy_threshold}. The energy threshold is improved by at least 5.1\% and by at most 22.3\%. The best improvements ($\ge$ 18.6\%) are observed for simulations taken at high NSB level (400 MHz), while modest improvements (5.1\% - 7.3\%) are seen for the low NSB level (75 MHz). For the NSB level of 200 MHz, improvements were obtained in the range of 14.4\% - 15.5\%. 

\begin{figure}
    \centering
    \includegraphics[width=0.7\hsize]{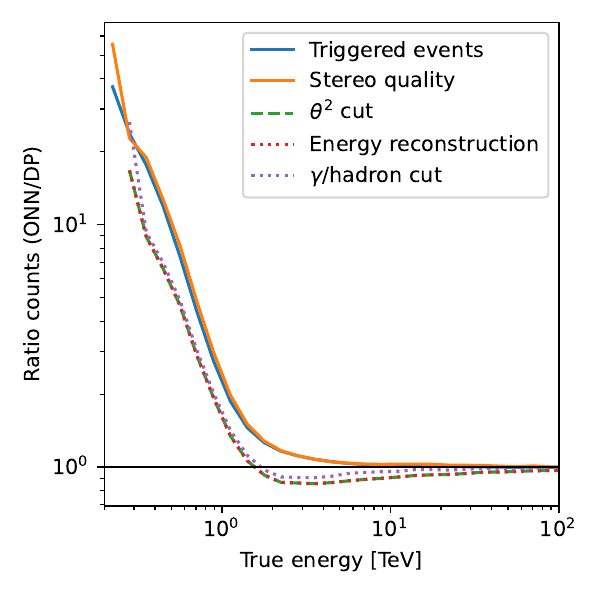}
    \caption{Ratio of number of counts (ONN by double-pass) after cuts applied for the effective area computation as a function of true energy.  Cuts are applied in consecutive order as they appear in the legend. Simulations correspond to $\phi_z$ = 60$^{\circ}$ and $\eta$ = 400 MHz.}
    \label{fig:efcuts}
\end{figure}

 Figures \ref{fig:engres} and \ref{fig:angres} give the comparisons for the energy and angular resolution, respectively, between the two cleaning methods. The difference between reconstructed and simulated shower energy, $E_{\mathrm{true}}$, is known as the energy bias: $\left( E_{\mathrm{rec}} - E_{\mathrm{true}} \right) / E_{\mathrm{true}}$, shown in Fig. \ref{fig:engbias}. The energy resolution is derived as the 68\% containment width around the mean value of the energy bias distribution. The angular resolution corresponds to the radius containing 68\% of simulated point source gamma rays~\citep{park2015performanceveritasexperiment}. Both quantities are dependent on the NSB level and on zenith and azimuth angles of observations.
As shown in the residual panels, the energy resolution for both methods agree within 10\% for all configuration above 1 TeV, where the resolution generally improves. The highest discrepancies are observed at large zenith and noise levels. For low and moderate zenith angles and below 1 TeV, the ONN method provides an improved energy resolution by up to $\sim$ 15\%, while a worsening of energy resolution is found by up to $\sim$ 6\%. Minor discrepancies are expected, particularly at low energies, where the ONN method benefits from an increased event sample. The angular resolution differs by up to $\sim$ 15\% above 1 TeV when considering all configurations. Below 1 TeV, a worsening by up to 20\% (40\%) was found for the low (high) zenith configurations. This behaviour is due to the larger number of pixels reconstructed per image with the ONN (shown in Fig. \ref{fig:npixevent}). While more pixels preserve additional shower information, they also result in less defined morphologies, which leads to a worse accuracy of the direction reconstruction.
The improvement in event retention at $\lesssim$ 100 GeV is directly related to a reduction in the energy bias. This is consistent with expectations since in this energy range, the event sample is no longer almost exclusively dominated by high-amplitude images for which $\left( E_{\mathrm{rec}} - E_{\mathrm{true}} \right) >> 0$. This results in a shift of the energy bias towards zero, as is desired in the ideal scenario.

When considering only the range $< 1$ TeV, effective areas with the ONN are lower by at most 5\% for all zenith angles. Variations above $\gtrsim 1$ TeV are mainly driven by limitations from the geometry of the array and can be attributed to statistical fluctuations of the simulations. On the other hand, variations at $\lesssim$ 3 TeV, which are mainly due to limitations from cleaning and triggering efficiency, can be attributed to the worsening of the angular resolution, as shown in Fig. \ref{fig:efcuts}, where we present the ratio of surviving triggered counts after the application of several quality cuts. Simulations for Fig. \ref{fig:efcuts} were produced for $\phi_z$ = 60$^{\circ}$ and and $\eta$ = 400 MHz, which show the lowest effective area ratio and highest angular resolution degradation. Namely, the cuts are applied in the consecutive order: a stereo quality cut ($\chi^2 > 0$), a direction ($\theta^2$) reconstruction cut, a quality cut for the energy reconstruction and $\gamma$/hadron selection cuts. According to results shown in the figure, the lower performance of the ONN is due to the $\theta^2$ cut, which aligns with the worsening of the angular reconstruction. In the case of supersoft cuts, this corresponds to a strict threshold  of $\theta^2 < 0.012$ deg$^{2}$. Since box $\gamma$/hadron cuts for this section were optimised with the double-pass method, the observed reduction in effective area remains relatively modest. The worst effects happen particularly at large NSB levels ($\gtrsim 200$ MHz) and zenith angles ($\gtrsim$ 40$^{\circ}$), which are already expected to have the lowest reconstruction efficiencies.

\section{Performance study with Crab Nebula data}
\label{sec:data_performance}

\begin{figure}
   \centering
   \includegraphics[width=0.97\hsize]{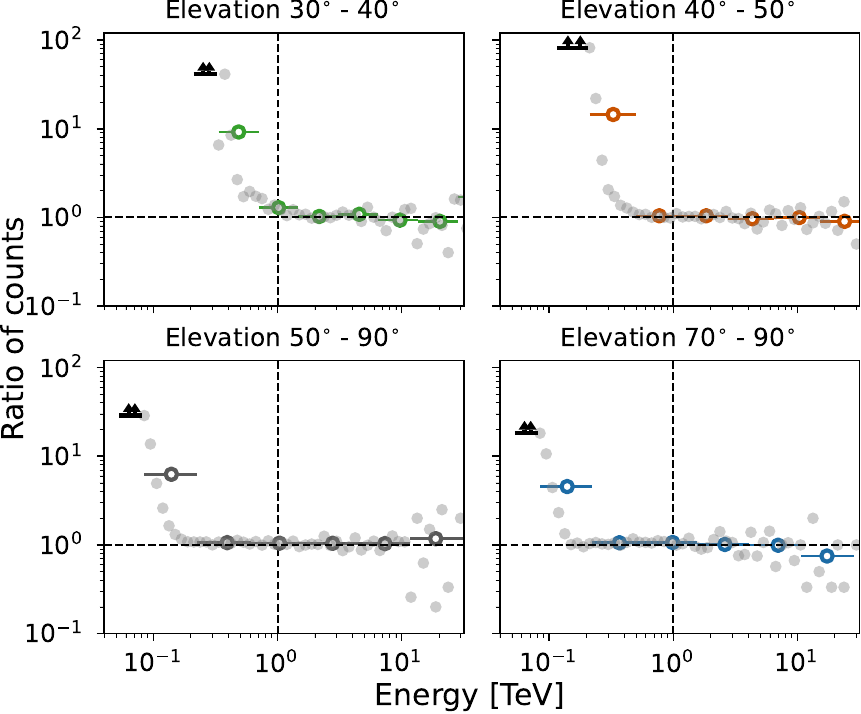}
     \caption{Ratio of the Crab Nebula excess counts reconstructed according to the ONN and double-pass cleaning methods for different configurations of the elevation angle. Ratios are only shown where possible, i.e. for energies with at least one excess event reconstructed per cleaning method. Ratios at energy ranges for which the ONN image cleaning reconstructs events while the nominal cleaning is completely suppressed are shown by the lower limits (black arrows). The ratio was also averaged over eight logarithmic bins (transparent circular markers) to help with visualisation.}
      \label{fig:ratio_excess_counts_oneplot}
\end{figure}

The performance of the ONN  image cleaning method was assessed with data collected for the Crab Nebula (review available in \citealt{crab}), which is widely regarded as a standard candle for VHE gamma rays and used as a reference for evaluating spectral features that may be introduced by new analysis techniques. The Crab Nebula dataset comprises a total of $\sim$ 121 hours of live-time collected between 2012 and 2018 and covers a wide range of elevation angles from 30$^\circ$ up to 90$^{\circ}$. The dataset undergoes the same analysis pipeline, differing only in the image cleaning method employed. In particular, after the cleaning, image quality cuts involve rejecting images which do not meet a minimum image brightness of 100 digital counts, at least 5 pixels per image and a loss parameter (the ratio between number of image pixels at the camera edge by the total number of pixels in the image) lower than 20\%. For this analysis, $\gamma$/hadron separation cuts were also optimised for supersoft sources. The background was estimated with the reflected regions technique \citep{Berge_2007}, assuming six background regions. To analyse the effects of the reduction of the energy threshold, the dataset is divided into the following ranges in elevation angles: [30$^\circ$ - 40$^{\circ}$], [40$^\circ$ - 50$^{\circ}$], [50$^\circ$ - 90$^{\circ}$] and [70$^\circ$ - 90$^{\circ}$]. The corresponding total live-time and analysis results for each elevation range can be found in Table \ref{tab:crabnebula}.

\begin{figure}
   \centering
   \includegraphics[width=0.97\hsize]{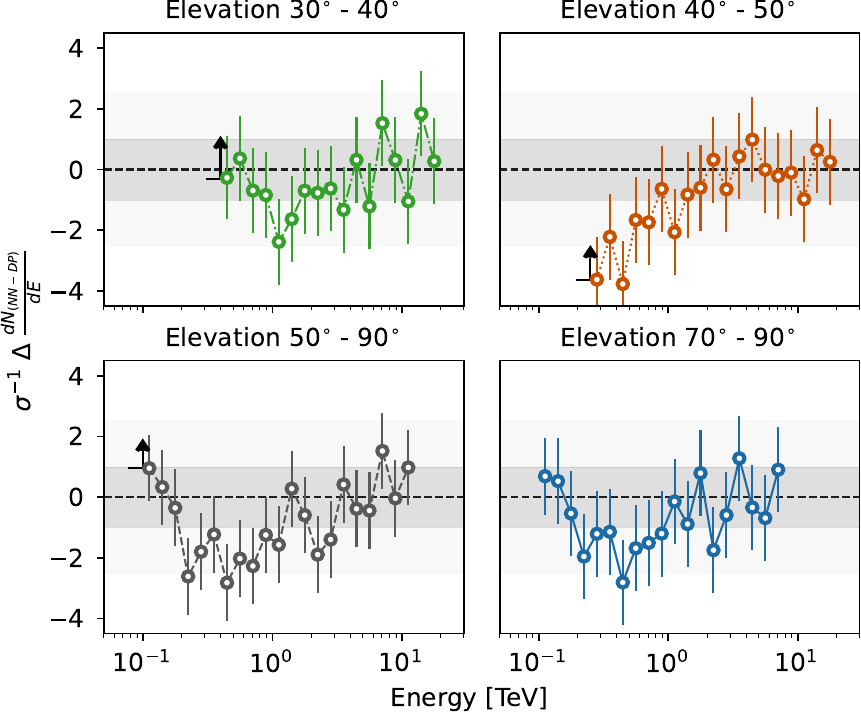}
     \caption{ Comparison of the Crab Nebula flux spectrum reconstructed according to the next neighbour and double-pass cleaning methods at different elevation ranges. Grey areas (from darker to lighter shade) indicate a difference of 1$\sigma$ and 2.5$\sigma$ between spectral points. Lower limits (black arrow) indicate a valid reconstructed spectral point with the ONN cleaning due the gain in excess events that have not been successfully reconstructed with the nominal method.}
      \label{fig:fit_crabnebula}
\end{figure}

Figure \ref{fig:ratio_excess_counts_oneplot} presents the ratio of excess counts from the Crab Nebula in the elevation configurations listed above. The ratio is given between the ONN and double-pass. We can generally conclude that a higher gamma-ray retention power is achieved with the ONN image cleaning in comparison to the nominal method, as seen from the increased counts spectra at sub-TeV energies. A greater number of events by factor of at most $\sim$ 41 (82, 29, 18) is seen below 1 TeV at elevations $\in$ [30$^\circ$, 40$^{\circ}$] ([40$^\circ$, 50$^{\circ}$], [50$^\circ$, 90$^{\circ}$], and [70$^\circ$, 90$^{\circ}$]). A total of 22 source events with energies as low as 63 GeV, which were suppressed after the nominal cleaning, are now successfully reconstructed for the elevation range [70$^\circ$, 90$^{\circ}$]. For comparison, with the nominal cleaning, the lowest reconstructed energy at this elevation range is 79 GeV. Similarly, at the \textit{INT\_El} range, the lowest reconstructed energy with the ONN method is of 141 GeV, against 199 GeV with the nominal cleaning. A total of 24 source counts are reconstructed at the newly accessible range. At the low elevation range (\textit{LOW\_El}), the lowest reconstructed energy is of 251 GeV with the ONN, against 316 GeV in nominal. A total of 31 new excess events are recovered.

As shown in Table \ref{tab:crabnebula}, an increase of 13-22\% is seen in the total reconstructed excess counts rate, along with an increase of 135-175\% in the reconstructed background rate. These results indicate an overall acquired efficacy in the event reconstruction power, whether for gamma-rays originating from the source or from the cosmic-ray background. The substantial increase in the background rate is anticipated due to the intensity and steeper spectrum of the cosmic-ray flux in comparison to the Crab Nebula gamma-ray flux. The substantial increase in the background rate results in a reduction in significance of $\sim$ 12\%, as observed among all configurations.%

Figure \ref{fig:fit_crabnebula} presents the comparison of the flux spectrum of the Crab Nebula between both image cleaning methods. Flux points are reconstructed with a $\Delta$log$_{10} \left( E_{\rm rec} ~/~ {\rm TeV}\right)$ = -0.2 energy binning and require at least 3$\sigma$ and 10 excess events per bin. The shaded areas (from darker to lighter gray) represent a difference in spectral flux points of 1$\sigma$ and 2.5$\sigma$, respectively. Most flux points differ by less than 2.3 standard deviations (with all flux points differing by, at most, 3.8$\sigma$ for the elevation range 40$^\circ$ - 50$^\circ$), indicating a good agreement for the spectral reconstruction between both methods. Due to the major gains in event reconstruction at low energies, the ONN method was able to reconstruct significant spectral points which were completely suppressed by the traditional image cleaning method. These spectral points are indicated by the lower limits in the corresponding energy bin. %

\begin{table*}
\centering
\caption{Parameters and analysis results for Crab Nebula datasets cleaned with the ONN method (in blue) and with the traditional cleaning method (in grey).}
\label{tab:crabnebula}
\begin{tabular}{cccccccc}
\hline
\hline
\begin{tabular}[c]{@{}c@{}}Elevation range\\ {[}deg{]}\end{tabular} & Label & \begin{tabular}[c]{@{}c@{}}Live-time\\ {[}hours{]}\end{tabular} & \begin{tabular}[c]{@{}c@{}}N$_\mathrm{excess}$\\ {[}$\times$ 10$^3${]}\end{tabular}                                  & \begin{tabular}[c]{@{}c@{}}N$_\mathrm{off}$\\ {[}$\times$ 10$^3${]}\end{tabular}                                     & \begin{tabular}[c]{@{}c@{}}Significance\\ {[}sigma{]}\end{tabular} & excess/min                                     & bkg/min                                       \\ \hline
30 - 40  & LOW\_El                                                            & 33.1                                                               & \begin{tabular}[c]{@{}c@{}} \textcolor{blue}{10.39}\\ \textcolor{gray}{8.48} \end{tabular} & \begin{tabular}[c]{@{}c@{}}\textcolor{blue}{5.80}\\ \textcolor{gray}{2.11}\end{tabular} & \begin{tabular}[c]{@{}c@{}}\textcolor{blue}{100}\\ \textcolor{gray}{114}\end{tabular}                      & \begin{tabular}[c]{@{}c@{}}\textcolor{blue}{5.23 $\pm$ 0.07}\\ \textcolor{gray}{4.27 $\pm$ 0.05}\end{tabular} & \begin{tabular}[c]{@{}c@{}}\textcolor{blue}{2.92}\\ \textcolor{gray}{1.06}\end{tabular} \\
40 - 50 &  INT\_El                                                             & 30.8                                                               & \begin{tabular}[c]{@{}c@{}}\textcolor{blue}{16.46}\\ \textcolor{gray}{13.74}\end{tabular} & \begin{tabular}[c]{@{}c@{}}\textcolor{blue}{8.31}\\ \textcolor{gray}{3.25} \end{tabular} & \begin{tabular}[c]{@{}c@{}}\textcolor{blue}{130}\\ \textcolor{gray}{147} \end{tabular}                      & \begin{tabular}[c]{@{}c@{}}\textcolor{blue}{8.90 $\pm$ 0.09}\\ \textcolor{gray}{7.42 $\pm$ 0.07}\end{tabular} & \begin{tabular}[c]{@{}c@{}}\textcolor{blue}{4.49}\\ \textcolor{gray}{1.76}\end{tabular} \\
50 - 90  &  INT\_El+                                                                  & 43.2                                                               & \begin{tabular}[c]{@{}c@{}}\textcolor{blue}{40.57}\\ \textcolor{gray}{34.09} \end{tabular} & \begin{tabular}[c]{@{}c@{}}\textcolor{blue}{19.80}\\ \textcolor{gray}{8.28} \end{tabular} & \begin{tabular}[c]{@{}c@{}}\textcolor{blue}{206}\\ \textcolor{gray}{231} \end{tabular}                      & \begin{tabular}[c]{@{}c@{}}\textcolor{blue}{15.7 $\pm$ 0.1}\\ \textcolor{gray}{13.15 $\pm$ 0.08}\end{tabular} & \begin{tabular}[c]{@{}c@{}}\textcolor{blue}{7.64}\\ \textcolor{gray}{3.19}\end{tabular} \\
70 - 90   &  HIGH\_El                                                                 & 13.6

&

\begin{tabular}[c]{@{}c@{}}\textcolor{blue}{13.58} \\ \textcolor{gray}{11.60}\end{tabular} 

& 

\begin{tabular}[c]{@{}c@{}}\textcolor{blue}{6.64}\\ \textcolor{gray}{2.82} \end{tabular} 

& 

\begin{tabular}[c]{@{}c@{}}\textcolor{blue}{119}\\ \textcolor{gray}{134} \end{tabular}

&

\begin{tabular}[c]{@{}c@{}}\textcolor{blue}{16.17 $\pm$ 0.2}\\ \textcolor{gray}{14.3 $\pm$ 0.2} \end{tabular} 

& 

\begin{tabular}[c]{@{}c@{}}\textcolor{blue}{8.16}\\ \textcolor{gray}{3.46} \end{tabular} \\ \hline
\end{tabular}
\tablefoot{Total dataset live-time, number of excess events (N$_\mathrm{excess}$), number of events in background regions (N$_\mathrm{off}$), dataset significance, excess-rate per minute (excess/min) and background rate per minute (bkg/min). Significances are calculated according to the Li\&Ma statistical method, i.e. Eq. 17 on Ref. \citep{liandma}. A label is given to each configuration according to the range of elevation: low elevation angles (LOW\_El), intermediate elevation angles (INT\_El), intermediate and high elevation angles (INT\_El+), and high elevation angles (HIGH\_El).}
\end{table*}

\subsection{Results with $\gamma$/hadron cuts generated with boosted decision trees}
\label{sec:bdt}

The separation of gamma-ray events from the cosmic-ray background can be accomplished using boosted decision trees (BDTs) algorithms, which combine Hillas image parameters into discriminating variables that indicate if an event is gamma-ray-like or cosmic-ray-like. BDT-based $\gamma$/hadron separation cuts were generated using the Multi-Variate Analysis (TMVA) package \citep{tmva} and are trained separately for each cleaning method. The BDTs are trained using simulated gamma-ray showers as signal and data from control regions (without known gamma-ray sources or bright stars) as background. The work of \citep{krauseImprovedGammaHadron2017} provides a detailed overview of the VERITAS BDT training procedure and evaluates its performance in comparison to traditional box-cut methods.

\begin{figure}
    \centering
    \includegraphics[width=0.97\hsize]{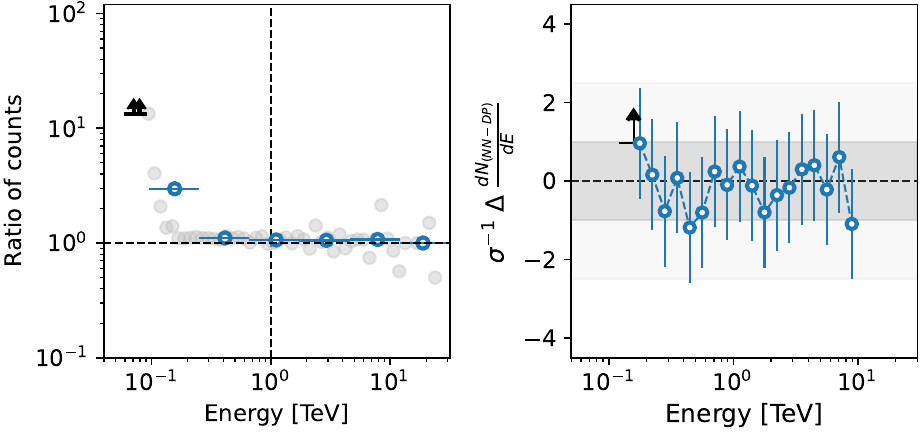}
     \caption{\textit{Left}: Ratio of excess counts spectrum for the Crab Nebula datasets cleaned with the ONN method and for the traditional cleaning method. Results are shown for $\gamma$/hadron separation cuts optimised with multivariate techniques. \textit{Right}: Comparison of the Crab Nebula flux spectrum reconstructed according to the ONN and double-pass cleaning methods. Grey areas (from darker to lighter shade) indicate a difference of 1$\sigma$ and 2.5$\sigma$.}
      \label{fig:crab_bdt}
\end{figure}

The Crab Nebula dataset consists of $\sim$ 20 hours of exposure-time with elevation angles ranging from $\sim$ 51 - 80 degrees. The dataset was cleaned with each different algorithm and the respective BDT $\gamma$/hadron cuts were applied. Similar to box cuts, we notice an increased retention of gamma-ray-like events, particularly at low-energies. With the new cuts, an increased significance of the dataset and excess event rate is achieved (see Table \ref{tab:crab_bdts}). Excess counts are increase by a factor of at most $\sim$ 14 below 1 TeV, as seen by Fig. \ref{fig:crab_bdt}. The minimum reconstructed energy with the ONN image cleaning is 71 GeV against 90 GeV with the nominal method.  A total of 57 excess events are gained in the additional energy range. Reconstructed flux points agree within both methods within at least $\sim$ 1.2$\sigma$, as is seen in Fig. \ref{fig:crab_bdt}

\begin{table*}[h!]
\centering
\caption{Parameters and analysis results for Crab Nebula datasets cleaned with the ONN method (in blue) and with the traditional cleaning method (in grey) and considering BDT cuts and supersoft box cuts for $\gamma$/hadron separation.}
\label{tab:crab_bdts}
\begin{tabular}{ccccccc}
\hline
\begin{tabular}[c]{@{}c@{}}Elevation range\\ {[}deg{]}\end{tabular} & \begin{tabular}[c]{@{}c@{}}live-time\\ {[}hours{]}\end{tabular} & \begin{tabular}[c]{@{}c@{}}N$_\mathrm{excess}$\\ {[}$\times$ 10$^3${]}\end{tabular}      & \begin{tabular}[c]{@{}c@{}}N$_\mathrm{off}$\\ {[}$\times$ 10$^3${]}\end{tabular}         & \begin{tabular}[c]{@{}c@{}}Significance\\ {[}sigma{]}\end{tabular}                         & excess/min                                                                                                   & bkg/min                                                   \\ \hline  
\multicolumn{7}{c}{BDT $\gamma$/hadron cuts}  \vspace{0.05cm} \\ \hline     
51 - 80                                                             & 19.8                                                            & \begin{tabular}[c]{@{}c@{}}\textcolor{blue}{15.5} \\ \textcolor{gray}{13.4}\end{tabular} & \begin{tabular}[c]{@{}c@{}}\textcolor{blue}{1.21} \\ \textcolor{gray}{0.75}\end{tabular} & \begin{tabular}[c]{@{}c@{}}\textcolor{blue}{194.6} \\ \textcolor{gray}{189.1}\end{tabular} & \begin{tabular}[c]{@{}c@{}}\textcolor{blue}{13.1 $\pm$ 0.1} \\ \textcolor{gray}{11.3 $\pm$ 0.1}\end{tabular} & \begin{tabular}[c]{@{}c@{}}\textcolor{blue}{1.02}\\ \textcolor{gray}{0.63}\end{tabular} \\ \hline
\multicolumn{7}{c}{Box $\gamma$/hadron cuts}  \vspace{0.05cm} \\ \hline
51 - 80                                                             & 19.8                                                            & \begin{tabular}[c]{@{}c@{}}\textcolor{blue}{17.7} \\ \textcolor{gray}{16.2}\end{tabular} & \begin{tabular}[c]{@{}c@{}}\textcolor{blue}{10.4} \\ \textcolor{gray}{4.6}\end{tabular}  & \begin{tabular}[c]{@{}c@{}}\textcolor{blue}{128.2} \\ \textcolor{gray}{152.8}\end{tabular} & \begin{tabular}[c]{@{}c@{}}\textcolor{blue}{14.9 $\pm$ 0.1} \\ \textcolor{gray}{13.6 $\pm$ 0.1}\end{tabular} & \begin{tabular}[c]{@{}c@{}}\textcolor{blue}{8.7}\\ \textcolor{gray}{3.8}\end{tabular}   \\ \hline
\end{tabular}
\tablefoot{Total dataset live-time, number of excess events (N$_\mathrm{excess}$), number of events in background regions (N$_\mathrm{off}$), dataset significance, excess-rate per minute (excess/min), and background rate per minute (bkg/min).}
\end{table*}

\subsection{Validation with Monte Carlo/data comparisons}

 The ONN was also validated by comparing the results of the image parametrisation of the Crab Nebula dataset with Monte Carlo simulations. This validation allows us to identify systematic uncertainties introduced by the cleaning and ensures consistent treatment of both real and simulated data. We performed the validation by comparing the distribution of the MSCW and MSCL parameters, which have been discussed in Sect. \ref{sec:racc}. %

\begin{figure}
    \centering
    {\includegraphics[width=0.97\hsize]{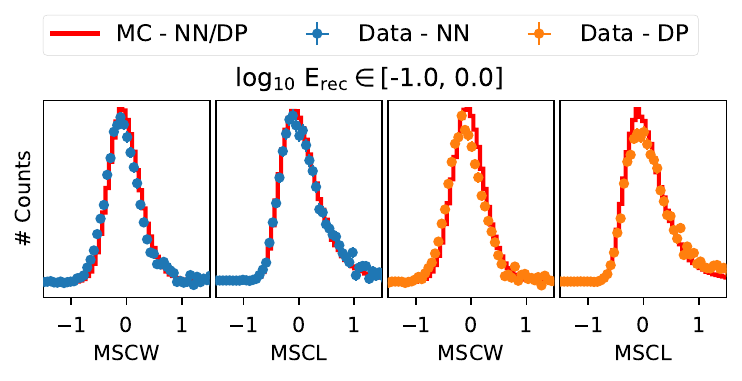}}
     \caption{Distribution of MSCW and MSCL shower parameters for Monte Carlo simulated events (red histograms) and the Crab Nebula dataset in blue (orange) for the ONN (double-pass, DP) cleaning. Parameter counts are arbitrarily normalised. Data and simulations are within the energy range of 100 GeV and 1 TeV.  Results up to 10 TeV and corresponding cumulative distributions are given in Fig. \ref{fig:full_mc_data_comparison}.}
      \label{fig:mccomparison_mscw}
\end{figure}

Figure \ref{fig:mccomparison_mscw} shows the distributions of reconstructed and simulated MSCW and MSCL between 100 GeV and 1 TeV for both cleaning methods. The data consists of excess events from $\sim$ 20 hours of live time recorded during the 2012–2013 observing season at elevation angles above 65 degrees. The simulated showers are generated at a zenith angle of 20 degrees with a noise level of 200 MHz. Figure \ref{fig:full_mc_data_comparison} shows comparisons for MSCW and MSCL in six energy bins between 100 GeV and 10 TeV together with the corresponding cumulative distributions. Overall, the Crab Nebula data and Monte Carlo simulations exhibit consistent behaviour, most clearly seen at low energies. Note that above $\sim$ 2 TeV, the reconstructed dataset has a limited event statistics, which makes comparisons less reliable. To quantify the level of agreement, a Kolmogorov–Smirnov (KS) test is performed on the cumulative distributions of the empirical and simulated samples. Below 1 TeV, the empirical distribution agrees with the simulated one within 2.0 $\sigma$ - 7.1 $\sigma$. It is important to note that discrepancies between data and simulations are expected, since they do not provide exact representations of real observations. For instance, simulations are generated at discrete zenith angles and noise levels and have a limited representation of factors such as telescope shadowing, atmospheric profiles, and hardware issues, such as broken channels. The limited representation of atmospheric profiles is known to introduce a systematic uncertainty of 10-15\% in the VERITAS analysis \citep{sfactorvts}. From the cumulative distribution of the stereo parameters, we derive a systematic uncertainty of 1.8-3.4\% below 1 TeV, which is well below other sources of systematic uncertainties. For the nominal cleaning, the agreement between the simulated and reconstructed distributions is at the level of 3.4 - 5.8 $\sigma$ at $<$ 1 TeV. Systematic uncertainties for the nominal cleaning are at a level of 3.2-5.4\% below 1 TeV.

\section{Differential sensitivity of the next neighbour image cleaning}
\label{sec:sensitivity}

\begin{figure}
    \centering
    {\includegraphics[width=0.97\hsize]{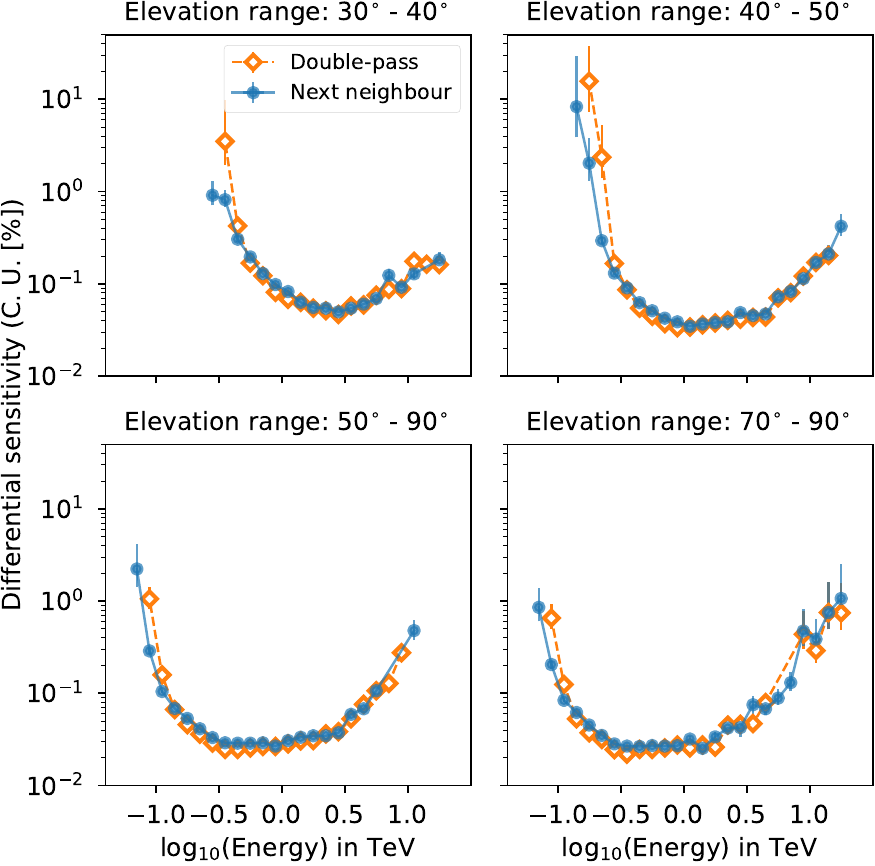}}
     \caption{VERITAS differential sensitivity for 50 hours of observations based on the ONN image cleaning method (blue) and for the double-pass cleaning method (orange).} %
      \label{fig:diffsen}
\end{figure}

\begin{figure}
   \centering
   \includegraphics[width=0.95\hsize]{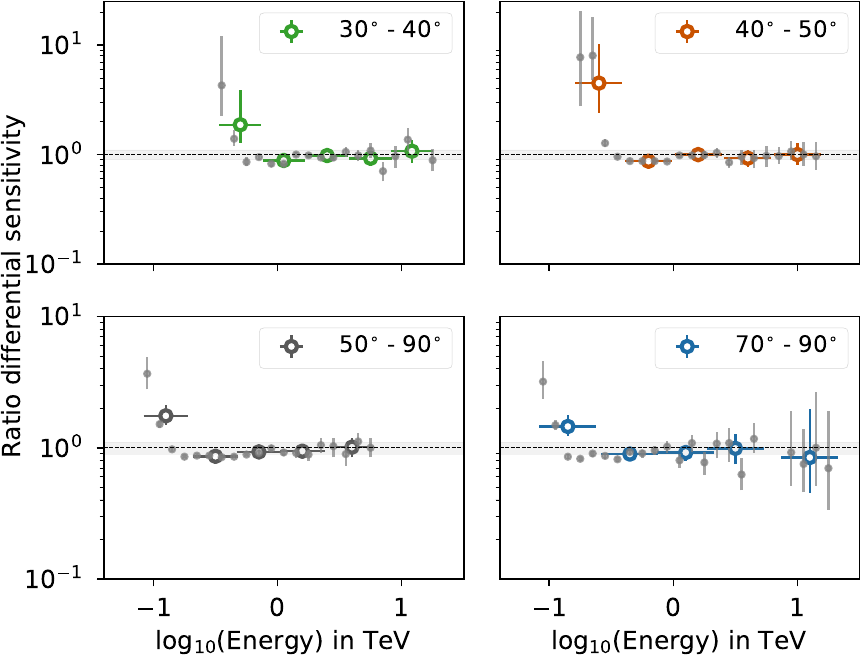}
      \caption{Ratio between the differential sensitivity of the nominal double-pass cleaning by the next neighbour method (DP/NN) in different elevation ranges. Here, a higher ratio indicates a better sensitivity of the ONN method. The ratio was also averaged over eight logarithmic bins (transparent circular markers) to help with visualisation.}
         \label{fig:ratiodiff}
\end{figure}

To provide a benchmark for the performance comparison, we provide in Fig. \ref{fig:diffsen} the differential sensitivity of VERITAS based on the same Crab Nebula dataset described in Table \ref{tab:crabnebula}. The differential sensitivity represents the energy flux that results in a significant detection after 50 hours of observations. The requirements for this calculation are a 5$\sigma$ significance and at least ten excess events per energy bin. Figure \ref{fig:ratiodiff} shows the ratio between methods, where applicable. A major improvement by a factor of 8.0 (4.3, 3.7 and 3.2) is seen for the first common energy bin for the INT\_El (LOW\_El, INT\_El+, HIGH\_El) configuration. As previously mentioned, the improvement in event retention is particularly important in the low-energy regime, which is seen by the extension of the differential sensitivity curves to energy levels completely suppressed by the nominal cleaning. We also note an overall minimum decrease in sensitivity by a factor of at 18\% (18\%, 14\%, 15\%) below 1 TeV for configuration LOW\_El (HIGH\_El, INT\_El, INT\_El+). This is due to the lower effective areas achieved in this regime (previously discussed for Fig. \ref{fig:effarea}). Ratios reported in this section are considered for $\lesssim$ 1 TeV due to statistical fluctuations of source events above this energy range.%

\section{Conclusions}
\label{sec:conclusions}

We present the application of the ONN image cleaning procedure for data collected by VERITAS, which efficiently uses the timing information of extensive air showers to decrease the pixel thresholds necessary to suppress the image noise. By lowering the pixel charge thresholds, shower images that do not meet the cleaning criteria imposed by the nominal cleaning (5$\sigma$/2.5$\sigma$) can be efficiently reconstructed, while still guaranteeing a low level of noise contamination. This mainly impacts low-energy showers, which commonly result in images of low overall brightness with fewer reconstructed pixels. The method had already been proven to improve event retention for CTAO simulations. In this study, we present the performance not only for Monte Carlo simulations of gamma-rays, but also for an extensive Crab Nebula dataset collected by VERITAS. Below, we summarise the results of the application of the ONN cleaning:

\paragraph{Results for the methodology:} the effect of the level of noise contamination was tested by studying the morphology of simulated gamma rays and the survival rate of clusters in pedestal images. An optimal range of $5\cdot 10^{-5} \lesssim f_{\mathrm{acc}} \lesssim 5\cdot 10^{-4}$ was found to offer the best balance between suppression of noise and retention of Cherenkov signals and image morphology.

 \paragraph{Results for the Monte Carlo simulations:} gamma-ray showers were simulated considering different zenith angles and noise level configurations. Overall, we noticed that the ONN method increases effective areas at low energies for all configurations. In addition, effective areas extend to energies inaccessible by the nominal cleaning due to the complete suppression of events during the cleaning process. As a direct consequence, energy thresholds were improved by a factor of at most 0.78. We see no large negative impact on the energy resolution, although the angular resolution shows a relatively worse performance with the new method across different zenith angles. This is expected due to the higher retention of pixels per image.

\paragraph{Results for the Crab Nebula:} a total of 120 hours of Crab Nebula data were used to assess the performance of the ONN method. The dataset was divided into elevation angles to comprehend the impact at low energies. We noticed an overall higher event retention at low energies. The higher event retention rate allowed for the reconstruction of flux points suppressed by the nominal cleaning. With the addition of BDT $\gamma$/hadron cuts, the significance of the dataset was raised, while still showing improvements in event retention and an additional significant spectral point for the Crab Nebula flux.

The increased event retention power results in an increase in the differential sensitivity of VERITAS by a factor of up to $\sim$ 8 in the first common energy bin. It is also notable that the sensitivity of the telescopes extends to lower energies with the application of the ONN method.

As demonstrated via a Monte Carlo and data comparison, no significant systematic effects were introduced by the new method. Any differences seen between both distributions can be explained on the basis of the statistical limitations of the data and constraints on simulation production.

The ONN image cleaning procedure is especially advantageous for faint and transient signals, which benefit from the lower energy threshold, particularly when we are reconstructing energy spectra. This includes probing short-duration transient signals, active galactic nuclei flares, emission from galaxy clusters, and the search for gamma-ray signals from dark matter annihilation. The method should be suitable for reconstructing any IACT data, including CTAO, MAGIC, and H.E.S.S., as long as the arrival timing of extensive air showers is available at a reasonable resolution in the nanosecond range.

\begin{acknowledgements} This research is supported by grants from the U.S. Department of Energy Office of Science, the U.S. National Science Foundation and the Smithsonian Institution, by NSERC in Canada, and by the Helmholtz Association in Germany. We acknowledge the excellent work of the technical support staff at the Fred Lawrence Whipple Observatory and at the collaborating institutions in the construction and operation of the instrument. The authors would like to thank the VERITAS collaboration for the available data. M. K. also acknowledges the International Helmholtz-Weizmann Research School for Multimessenger Astronomy.
\end{acknowledgements}

\bibpunct{(}{)}{;}{a}{}{,}
\bibliographystyle{aa} 
\bibliography{bibliography}

@article{abe2023crab,
doi = {10.3847/1538-4357/ace89d},
url = {https://doi.org/10.3847/1538-4357/ace89d},
year = {2023},
month = {oct},
publisher = {\apj},
volume = {956},
number = {2},
pages = {80},
author = {Abe, H. and Abe, K. and Abe, S. and Aguasca-Cabot, A. and Agudo, I. and Alvarez Crespo, N. and Antonelli, L. A. and Aramo, C. and Arbet-Engels, A. and Arcaro, C. and Artero, M. and Asano, K. and Aubert, P. and Baktash, A. and Bamba, A. and Baquero Larriva, A. and Baroncelli, L. and Barres de Almeida, U. and Barrio, J. A. and Batkovic, I. and Baxter, J. and Becerra González, J. and Bernardini, E. and Bernardos, M. I. and Bernete Medrano, J. and Berti, A. and Bhattacharjee, P. and Biederbeck, N. and Bigongiari, C. and Bissaldi, E. and Blanch, O. and Bonnoli, G. and Bordas, P. and Borghese, A. and Bulgarelli, A. and Burelli, I. and Buscemi, M. and Cardillo, M. and Caroff, S. and Carosi, A. and Cassol, F. and Cauz, D. and Ceribella, G. and Chai, Y. and Cheng, K. and Chiavassa, A. and Chikawa, M. and Chytka, L. and Cifuentes, A. and Contreras, J. L. and Cortina, J. and Costantini, H. and D’Amico, G. and Dalchenko, M. and De Angelis, A. and de Bony de Lavergne, M. and De Lotto, B. and de Menezes, R. and Deleglise, G. and Delgado, C. and Delgado Mengual, J. and della Volpe, D. and Dellaiera, M. and Depaoli, D. and Di Piano, A. and Di Pierro, F. and Di Tria, R. and Di Venere, L. and Díaz, C. and Dominik, R. M. and Dominis Prester, D. and Donini, A. and Dorner, D. and Doro, M. and Elsässer, D. and Emery, G. and Escudero, J. and Fallah Ramazani, V. and Ferrara, G. and Ferrarotto, F. and Fiasson, A. and Freixas Coromina, L. and Fröse, S. and Fukami, S. and Fukazawa, Y. and Garcia, E. and Garcia López, R. and Gasbarra, C. and Gasparrini, D. and Geyer, F. and Giesbrecht Paiva, J. and Giglietto, N. and Giordano, F. and Giro, E. and Gliwny, P. and Godinovic, N. and Grau, R. and Green, D. and Green, J. and Gunji, S. and Hackfeld, J. and Hadasch, D. and Hahn, A. and Hashiyama, K. and Hassan, T. and Hayashi, K. and Heckmann, L. and Heller, M. and Herrera Llorente, J. and Hirotani, K. and Hoffmann, D. and Horns, D. and Houles, J. and Hrabovsky, M. and Hrupec, D. and Hui, D. and Hütten, M. and Iarlori, M. and Imazawa, R. and Inada, T. and Inome, Y. and Ioka, K. and Iori, M. and Ishio, K. and Iwamura, Y. and Jacquemont, M. and Jimenez Martinez, I. and Jurysek, J. and Kagaya, M. and Karas, V. and Katagiri, H. and Kataoka, J. and Kerszberg, D. and Kobayashi, Y. and Kong, A. and Kubo, H. and Kushida, J. and Lainez, M. and Lamanna, G. and Lamastra, A. and Le Flour, T. and Linhoff, M. and Longo, F. and López-Coto, R. and López-Moya, M. and López-Oramas, A. and Loporchio, S. and Lorini, A. and Luque-Escamilla, P. L. and Majumdar, P. and Makariev, M. and Mandat, D. and Manganaro, M. and Manicò, G. and Mannheim, K. and Mariotti, M. and Marquez, P. and Marsella, G. and Martí, J. and Martinez, O. and Martínez, G. and Martínez, M. and Marusevec, P. and Mas-Aguilar, A. and Maurin, G. and Mazin, D. and Mestre Guillen, E. and Micanovic, S. and Miceli, D. and Miener, T. and Miranda, J. M. and Mirzoyan, R. and Mizuno, T. and Molero Gonzalez, M. and Molina, E. and Montaruli, T. and Monteiro, I. and Moralejo, A. and Morcuende, D. and Morselli, A. and Mrakovcic, K. and Murase, K. and Nagai, A. and Nagataki, S. and Nakamori, T. and Nickel, L. and Nievas, M. and Nishijima, K. and Noda, K. and Nosek, D. and Nozaki, S. and Ohishi, M. and Ohtani, Y. and Oka, T. and Okazaki, N. and Okumura, A. and Orito, R. and Otero-Santos, J. and Palatiello, M. and Paneque, D. and Pantaleo, F. R. and Paoletti, R. and Paredes, J. M. and Pech, M. and Pecimotika, M. and Peresano, M. and Pérez, A. and Pietropaolo, E. and Pirola, G. and Plard, C. and Podobnik, F. and Poireau, V. and Polo, M. and Pons, E. and Prandini, E. and Prast, J. and Principe, G. and Priyadarshi, C. and Prouza, M. and Rando, R. and Rhode, W. and Ribó, M. and Rizi, V. and Rodriguez Fernandez, G. and Ruiz, J. E. and Saito, T. and Sakurai, S. and Sanchez, D. A. and Šarić, T. and Sato, Y. and Saturni, F. G. and Schleicher, B. and Schmuckermaier, F. and Schubert, J. L. and Schussler, F. and Schweizer, T. and Seglar Arroyo, M. and Silvia, R. and Sitarek, J. and Sliusar, V. and Spolon, A. and Strišković, J. and Strzys, M. and Suda, Y. and Sunada, Y. and Tajima, H. and Takahashi, H. and Takahashi, M. and Takata, J. and Takeishi, R. and Tam, P. H. T. and Tanaka, S. J. and Tateishi, D. and Tejedor, L. A. and Temnikov, P. and Terada, Y. and Terauchi, K. and Terzic, T. and Teshima, M. and Tluczykont, M. and Tokanai, F. and Torres, D. F. and Travnicek, P. and Truzzi, S. and Tutone, A. and Uhlrich, G. and Vacula, M. and Vallania, P. and van Scherpenberg, J. and Vázquez Acosta, M. and Verguilov, V. and Viale, I. and Vigliano, A. and Vigorito, C. F. and Vitale, V. and Voutsinas, G. and Vovk, I. and Vuillaume, T. and Walter, R. and Will, M. and Yamamoto, T. and Yamazaki, R. and Yoshida, T. and Yoshikoshi, T. and Zywucka, N. and Bernlöhr, K. and Gueta, O. and Kosack, K. and Maier, G. and Watson, J.},
title = {Observations of the Crab Nebula and Pulsar with the Large-sized Telescope Prototype of the Cherenkov Telescope Array},
journal = {The Astrophysical Journal}
}

@article{sfactorvts,
   title={The throughput calibration of the VERITAS telescopes},
   volume={658},
   ISSN={1432-0746},
   url={http://dx.doi.org/10.1051/0004-6361/202142275},
   DOI={10.1051/0004-6361/202142275},
   journal={\aap},
   publisher={EDP Sciences},
   author={Adams, C. B. and Benbow, W. and Brill, A. and Buckley, J. H. and Christiansen, J. L. and Falcone, A. and Feng, Q. and Finley, J. P. and Foote, J. and Fortson, L. and Furniss, A. and Giuri, C. and Hanna, D. and Hassan, T. and Hervet, O. and Holder, J. and Hona, B. and Humensky, T. B. and Jin, W. and Kaaret, P. and Kleiner, T. K. and Kumar, S. and Lang, M. J. and Lundy, M. and Maier, G. and Moriarty, P. and Mukherjee, R. and Nievas Rosillo, M. and O’Brien, S. and Park, N. and Patel, S. and Pfrang, K. and Pohl, M. and Prado, R. R. and Pueschel, E. and Quinn, J. and Ragan, K. and Reynolds, P. T. and Ribeiro, D. and Roache, E. and Ryan, J. L. and Santander, M. and Weinstein, A. and Williams, D. A. and Williamson, T. J.},
   year={2022},
   month=Feb, pages={A83} }

@BOOK{Aharonian_2004,
       author = {{Aharonian}, Felix A.},
        title = "{Very high energy cosmic gamma radiation : a crucial window on the extreme Universe}",
         year = 2004,
          doi = {10.1142/4657},
       publisher = {World Scientific},
       doi = {https://doi.org/10.1142/4657}
}

@article{corsika,
    author = "Alameddine, Jean-Marco and others",
    collaboration = "CORSIKA",
    title = "{The particle-shower simulation code CORSIKA 8}",
    eprint = "2308.05475",
    archivePrefix = "arXiv",
    primaryClass = "astro-ph.HE",
    doi = "10.22323/1.444.0310",
    journal = "PoS",
    volume = "ICRC2023",
    pages = "310",
    year = "2023"
}

@Article{crab,
AUTHOR = {Amato, Elena and Olmi, Barbara},
TITLE = {The Crab Pulsar and Nebula as Seen in Gamma-Rays},
JOURNAL = {Universe},
VOLUME = {7},
YEAR = {2021},
NUMBER = {11},
ARTICLE-NUMBER = {448},
URL = {https://www.mdpi.com/2218-1997/7/11/448},
ISSN = {2218-1997},
DOI = {10.3390/universe7110448}
}

@article{Berge_2007,
   title={Background modelling in very-high-energy <i>γ</i>-ray astronomy},
   volume={466},
   ISSN={1432-0746},
   url={http://dx.doi.org/10.1051/0004-6361:20066674},
   DOI={10.1051/0004-6361:20066674},
   number={3},
   journal={\aap},
   publisher={EDP Sciences},
   author={Berge, D. and Funk, S. and Hinton, J.},
   year={2007},
   month=Apr, pages={1219–1229} }

@article{hessic,
   author = {\'{C}elić, J. and Lang, R. G. and Steinmassl, S. and Hinton, J. and Funk, S.},
	title = {A novel approach to optimizing the image cleaning performance of Imaging Atmospheric Cherenkov Telescopes: Application to a time-based cleaning for H.E.S.S.},
	DOI= "10.1051/0004-6361/202554824",
	url= "https://doi.org/10.1051/0004-6361/202554824",
	journal = {A\&A},
	year = 2025,
	volume = 699,
	pages = "A96",
}

@article{ped,
   title={The VERITAS standard data analysis}, 
      author={M. K. Daniel},
      year={2007},
      journal = {ArXiv e-prints},
      archivePrefix = "arXiv",
      eprint={0709.4006},
      primaryClass={astro-ph},
      url={https://arxiv.org/abs/0709.4006}, 
}

@article{denauroisAnalysisMethodsAtmospheric2006,
      title={Analysis methods for Atmospheric Cerenkov Telescopes}, 
      author={Mathieu de Naurois},
      year={2006},
      booktitle = "{7th Workshop on Towards a Network of Atmospheric Cherenkov Detectors}",
      journal = {ArXiv e-prints},
archivePrefix = "arXiv",
      eprint={astro-ph/0607247},
      archivePrefix={arXiv},
      primaryClass={astro-ph},
      url={https://arxiv.org/abs/astro-ph/0607247}, 
}

@article{Fink:2015fia,
title = {PMT module characterization test bench for the MAGIC IACT telescopes},
journal = {Nucl. Instrum. Methods Phys. Res. A},
volume = {787},
pages = {250-253},
year = {2015},
issn = {0168-9002},
doi = {https://doi.org/10.1016/j.nima.2014.12.043},
url = {https://www.sciencedirect.com/science/article/pii/S0168900214014855},
author = {David Fink and Jürgen Hose and Olaf Reimann and David Paneque and Razmik Mirzoyan and Daisuke Nakajima and Koji Saito}
}

@article{Garczarczyk2011zz,
title = {The major atmospheric gamma-ray imaging Cherenkov telescope},
journal = {Nucl. Instrum. Methods Phys. Res. A},
volume = {639},
number = {1},
pages = {33-36},
year = {2011},
note = {Proceedings of the Seventh International Workshop on Ring Imaging Cherenkov Detectors},
issn = {0168-9002},
doi = {https://doi.org/10.1016/j.nima.2010.09.020},
url = {https://www.sciencedirect.com/science/article/pii/S0168900210020139},
author = {Markus Garczarczyk}
}

@article{sensi,
    author = "Gueta, Orel",
    collaboration = "CTA Consortium, CTA Observatory",
    title = "{The Cherenkov Telescope Array: layout, design and performance}",
    eprint = "2108.04512",
    archivePrefix = "arXiv",
    primaryClass = "astro-ph.IM",
    doi = "10.22323/1.395.0885",
    journal = "PoS",
    volume = "ICRC2021",
    pages = "885",
    year = "2021"
}

@INPROCEEDINGS{Hillas1985,
       author = {{Hillas}, A.~M.},
        title = "{Cerenkov Light Images of EAS Produced by Primary Gamma Rays and by Nuclei}",
    booktitle = {19th International Cosmic Ray Conference (ICRC19), Volume 3},
         year = 1985,
       editor = {{Jones}, F.~C.},
       series = {International Cosmic Ray Conference},
       volume = {3},
        month = aug,
        pages = {445},
       adsurl = {https://ui.adsabs.harvard.edu/abs/1985ICRC....3..445H}
}

@article{2013APh4319H,
title = {Evolution of ground-based gamma-ray astronomy from the early days to the Cherenkov Telescope Arrays},
journal = {Astroparticle Physics},
volume = {43},
pages = {19-43},
year = {2013},
note = {Seeing the High-Energy Universe with the Cherenkov Telescope Array - The Science Explored with the CTA},
issn = {0927-6505},
doi = {https://doi.org/10.1016/j.astropartphys.2012.06.002},
url = {https://www.sciencedirect.com/science/article/pii/S0927650512001326},
author = {A.M. Hillas}
}

@article{Hinton_2004,
title = {The status of the HESS project},
journal = {New Astronomy Reviews},
volume = {48},
number = {5},
pages = {331-337},
year = {2004},
note = {2nd VERITAS Symposium on the Astrophysics of Extragalactic Sources},
issn = {1387-6473},
doi = {https://doi.org/10.1016/j.newar.2003.12.004},
url = {https://www.sciencedirect.com/science/article/pii/S1387647303003555},
author = {J.A Hinton}
}

@techreport{tmva,
      author        = "Hocker, Andreas and Speckmayer, Peter and Stelzer, Jorg
                       and Therhaag, Jan and von Toerne, Eckhard and Voss, Helge
                       and Backes, Moritz and Carli, Tancredi and Cohen, Or and
                       Christov, Asen and Dannheim, Domikik and Danielowski,
                       Krzysztof and Henrot-Versille, S. and Jachowski, M. and
                       Kraszewski, Kamil and Krasznahorkay, A., Jr. and Kruk,
                       Maciej and Mahalalel, Y. and Ospanov, Rustem and Prudent,
                       X. and Robert, Arnaud and Schouten, Doug and Tegenfeldt, F.
                       and Voight, Alexander and Voss, K. and Wolter, Marcin and
                       Zemla, Andrzej",
      title         = "{TMVA - Toolkit for Multivariate Data Analysis with ROOT}",
      institution   = "CERN",
      reportNumber  = "physics/0703039, CERN-OPEN-2007-007",
      address       = "Geneva",
      year          = "2007",
      url           = "https://cds.cern.ch/record/1019880",
}

@article{Hofmann_1999,
title = {Comparison of techniques to reconstruct VHE gamma-ray showers from multiple stereoscopic Cherenkov images},
journal = {Astroparticle Physics},
volume = {12},
number = {3},
pages = {135-143},
year = {1999},
issn = {0927-6505},
doi = {https://doi.org/10.1016/S0927-6505(99)00084-5},
url = {https://www.sciencedirect.com/science/article/pii/S0927650599000845},
author = {W. Hofmann and I. Jung and A. Konopelko and H. Krawczynski and H. Lampeitl and G. Pühlhofer}
}

@article{doublepass,
author = {Holder, J.},
year = {2005},
month = {08},
pages = {},
title = "{Exploiting VERITAS timing information}",
volume = {5},
journal = {29th International Cosmic Ray Conference, ICRC 2005}
}

@article{Holder_2006,
   title="{The first VERITAS telescope}",
   volume={25},
   ISSN={0927-6505},
   url={http://dx.doi.org/10.1016/j.astropartphys.2006.04.002},
   DOI={10.1016/j.astropartphys.2006.04.002},
   number={6},
   journal={Astroparticle Physics},
   publisher={Elsevier BV},
   author={Holder, J. and Atkins, R.W. and Badran, H.M. and Blaylock, G. and Bradbury, S.M. and Buckley, J.H. and Byrum, K.L. and Carter-Lewis, D.A. and Celik, O. and Chow, Y.C.K. and Cogan, P. and Cui, W. and Daniel, M.K. and de la Calle Perez, I. and Dowdall, C. and Dowkontt, P. and Duke, C. and Falcone, A.D. and Fegan, S.J. and Finley, J.P. and Fortin, P. and Fortson, L.F. and Gibbs, K. and Gillanders, G. and Glidewell, O.J. and Grube, J. and Gutierrez, K.J. and Gyuk, G. and Hall, J. and Hanna, D. and Hays, E. and Horan, D. and Hughes, S.B. and Humensky, T.B. and Imran, A. and Jung, I. and Kaaret, P. and Kenny, G.E. and Kieda, D. and Kildea, J. and Knapp, J. and Krawczynski, H. and Krennrich, F. and Lang, M.J. and LeBohec, S. and Linton, E. and Little, E.K. and Maier, G. and Manseri, H. and Milovanovic, A. and Moriarty, P. and Mukherjee, R. and Ogden, P.A. and Ong, R.A. and Petry, D. and Perkins, J.S. and Pizlo, F. and Pohl, M. and Quinn, J. and Ragan, K. and Reynolds, P.T. and Roache, E.T. and Rose, H.J. and Schroedter, M. and Sembroski, G.H. and Sleege, G. and Steele, D. and Swordy, S.P. and Syson, A. and Toner, J.A. and Valcarcel, L. and Vassiliev, V.V. and Wakely, S.P. and Weekes, T.C. and White, R.J. and Williams, D.A. and Wagner, R.},
   year={2006},
   month=jul, pages={391–401} }

@ARTICLE{kieda2013gamma,
      title="{The Gamma Ray Detection sensitivity of the upgraded VERITAS Observatory}", 
      author={D. B. Kieda},
      year={2013},
      journal = {ArXiv e-prints},
      archivePrefix = "arXiv",
      eprint={1308.4849},
      archivePrefix={arXiv},
      primaryClass={astro-ph.IM},
      url={https://arxiv.org/abs/1308.4849}, 
}

@article{krauseImprovedGammaHadron2017,
  title = {Improved $\gamma$/Hadron Separation for the Detection of Faint Gamma-Ray Sources Using Boosted Decision Trees},
  author = {Krause, Maria and Pueschel, Elisa and Maier, Gernot},
  year = {2017},
  month = mar,
  journal = {Astroparticle Physics},
  volume = {89},
  eprint = {1701.06928},
  primaryclass = {astro-ph},
  pages = {1--9},
  issn = {09276505},
  doi = {10.1016/j.astropartphys.2017.01.004},
  urldate = {2024-01-21},
  archiveprefix = {arxiv},
  langid = {english}
}

@ARTICLE{nsb,
       author = {{Leinert}, Ch. and {Bowyer}, S. and {Haikala}, L.~K. and {Hanner}, M.~S. and {Hauser}, M.~G. and {Levasseur-Regourd}, A. -Ch. and {Mann}, I. and {Mattila}, K. and {Reach}, W.~T. and {Schlosser}, W. and {Staude}, H.~J. and {Toller}, G.~N. and {Weiland}, J.~L. and {Weinberg}, J.~L. and {Witt}, A.~N.},
        title = "{The 1997 reference of diffuse night sky brightness}",
      journal = {\aaps},
         year = 1998,
        month = jan,
       volume = {127},
        pages = {1-99},
          doi = {10.1051/aas:1998105},
       adsurl = {https://ui.adsabs.harvard.edu/abs/1998A&AS..127....1L}
}

@article{Lessard_2001,
title = {A new analysis method for reconstructing the arrival direction of TeV gamma rays using a single imaging atmospheric Cherenkov telescope},
journal = {Astroparticle Physics},
volume = {15},
number = {1},
pages = {1-18},
year = {2001},
issn = {0927-6505},
doi = {https://doi.org/10.1016/S0927-6505(00)00133-X},
url = {https://www.sciencedirect.com/science/article/pii/S092765050000133X},
author = {R.W. Lessard and J.H. Buckley and V. Connaughton and S. {Le Bohec}}
}

@misc{liandma,
 author = {Li, T. P. and Ma, Y. Q.},
 title = "{Analysis methods for results in gamma-ray astronomy.}",
 journal = {\apj},
 year = 1983,
 month = sep,
 volume = {272},
 pages = {317-324},
 doi = {10.1086/161295},
 adsurl = {https://ui.adsabs.harvard.edu/abs/1983ApJ...272..317L}
 }

@article{ctapipe-icrc-2023,
       author = {{Linhoff, M., Beiske, L., Biederbeck, N., et al. for the CTA Consortium and Observatory}},
        title = "{ctapipe - Prototype Open Event Reconstruction Pipeline for the Cherenkov Telescope Array}",
  doi = "10.22323/1.444.0703",
  journal = "PoS",
  year = 2023,
  volume = "ICRC2023",
  pages = "703"
}

@INPROCEEDINGS{lstchain_adass_2020,
       author = {{L{\'o}pez-Coto}, Ruben and {Baquero}, A. and {Bernardos}, M.~I. and {Cassol}, F. and {Foffano}, L. and {Garc{\'\i}a}, E. and {Gliwny}, P. and {Iwamura}, Y. and {Jacquemont}, M. and {Jouvin}, L. and {Kobayashi}, Y. and {Moralejo}, A. and {Morcuende}, D. and {Neise}, D. and {Nozaki}, S. and {N{\"o}the}, M. and {Priyadarshi}, C. and {Renier}, Y. and {Saha}, L. and {Sakurai}, S. and {Sitarek}, J. and {Takahashi}, M.},
        title = "{lstchain: An Analysis Pipeline for LST-1, the First Prototype Large-Sized Telescope of CTA}",
    booktitle = {Astronomical Data Analysis Software and Systems XXX},
         year = 2022,
       editor = {{Ruiz}, Jose Enrique and {Pierfedereci}, Francesco and {Teuben}, Peter},
       series = {Astronomical Society of the Pacific Conference Series},
       volume = {532},
        month = jul,
        pages = {357},
       adsurl = {https://ui.adsabs.harvard.edu/abs/2022ASPC..532..357L}
}

@article{maierEventdisplayAnalysisReconstruction2017,
  author = "Maier, Gernot  and  Holder, Jamie",
  title = "{Eventdisplay: An Analysis and Reconstruction Package for Ground-based Gamma-ray Astronomy}",
  doi = "10.22323/1.301.0747",
  journal = "PoS",
  year = 2017,
  volume = "ICRC2017",
  pages = "747"
}

@inproceedings{vtspmt,
    author = "Otte, A. Nepomuk and Gebremedhin, L. and Kaplan, K. and Long, D.",
    collaboration = "VERITAS",
    title = "{The Upgrade of VERITAS with High Efficiency Photomultipliers}",
    booktitle = "{32nd International Cosmic Ray Conference}",
    eprint = "1110.4702",
    archivePrefix = "arXiv",
    primaryClass = "astro-ph.IM",
    doi = "10.7529/ICRC2011/V09/1305",
    volume = "9",
    pages = "247",
    month = "10",
    year = "2011"
}

@article{park2015performanceveritasexperiment,
  author = "Park, Nahee",
  title = "{Performance of the VERITAS experiment }",
  doi = "10.22323/1.236.0771",
  journal = "PoS",
  year = 2016,
  volume = "ICRC2015",
  pages = "771"
}

@misc{perkins2009veritas,
      title={VERITAS Telescope 1 Relocation: Details and Improvements}, 
      author={Jeremy S. Perkins and Gernot Maier and {The VERITAS Collaboration}},
      year={2009},
      eprint={0912.3841},
      archivePrefix={arXiv},
      primaryClass={astro-ph.IM},
      url={https://arxiv.org/abs/0912.3841}, 
}

@article{Preu_2002,
   title={Study of the photon flux from the night sky at La Palma and Namibia, in the wavelength region relevant for imaging atmospheric Cherenkov telescopes},
   volume={481},
   ISSN={0168-9002},
   url={http://dx.doi.org/10.1016/S0168-9002(01)01264-5},
   DOI={10.1016/s0168-9002(01)01264-5},
   number={1–3},
   journal={Nuclear Instruments and Methods in Physics Research Section A},
   publisher={Elsevier BV},
   author={Preuß, S. and Hermann, G. and Hofmann, W. and Kohnle, A.},
   year={2002},
   month=apr, pages={229–240} }

@article{Sajjad,
doi = {10.1088/1367-2630/acecd9},
url = {https://doi.org/10.1088/1367-2630/acecd9},
year = {2023},
month = {aug},
publisher = {IOP Publishing},
volume = {25},
number = {8},
pages = {083038},
author = {Sajjad, Saeeda},
title = {The effect of night sky background on source reconstruction in atmospheric Cherenkov telescope arrays and implications for image cleaning},
journal = {New Journal of Physics}
}

@INPROCEEDINGS{shaydukNewImageCleaning2005,
       author = {{Shayduk}, M. and {Hengstebeck}, T. and {Kalekin}, O. and {Pavel}, N.~A. and {Schweizer}, T.},
        title = "{A New Image Cleaning Method for the MAGIC Telescope}",
    booktitle = {29th International Cosmic Ray Conference (ICRC29), Volume 5},
         year = 2005,
       series = {International Cosmic Ray Conference},
       volume = {5},
        month = jan,
        pages = {223},
       adsurl = {https://ui.adsabs.harvard.edu/abs/2005ICRC....5..223S}
}

@INPROCEEDINGS{onn,
       author = {{Shayduk}, M. and {CTA Consortium}},
        title = "{Optimized Next-neighbour Image Cleaning Method for Cherenkov Telescopes}",
    booktitle = {International Cosmic Ray Conference},
         year = 2013,
       series = {International Cosmic Ray Conference},
       volume = {33},
        month = jan,
        pages = {3000},
          doi = {10.48550/arXiv.1307.4939},
archivePrefix = {arXiv},
       eprint = {1307.4939},
 primaryClass = {astro-ph.IM},
       adsurl = {https://ui.adsabs.harvard.edu/abs/2013ICRC...33.3000S}
}

@misc{Weekes:2005da,
      title={The Atmospheric Cherenkov Imaging Technique for Very High Energy Gamma-ray Astronomy}, 
      author={Trevor C. Weekes},
      year={2005},
      eprint={astro-ph/0508253},
      archivePrefix={arXiv},
      primaryClass={astro-ph},
      url={https://arxiv.org/abs/astro-ph/0508253}, 
}

@INPROCEEDINGS{trigger,
       author = {{Weinstein}, A.},
        title = "{The VERITAS Trigger System}",
    booktitle = {International Cosmic Ray Conference},
         year = 2008,
       series = {International Cosmic Ray Conference},
       volume = {3},
        month = jan,
        pages = {1539-1542},
          doi = {10.48550/arXiv.0709.4438},
archivePrefix = {arXiv},
       eprint = {0709.4438},
 primaryClass = {astro-ph},
       adsurl = {https://ui.adsabs.harvard.edu/abs/2008ICRC....3.1539W}
}

@article{zorn2019sensitivityimprovementsveryhighenergygammaray,
  author = "Zorn, Justus",
  title = "{Sensitivity Improvements of Very-High-Energy $\gamma$-Ray Detection with the Upgraded H.E.S.S. I Cameras using Full Waveform Processing}",
  doi = "10.22323/1.358.0834",
  journal = "PoS",
  year = 2019,
  volume = "ICRC2019",
  pages = "834"
}

\begin{appendix}
\section{Contours for different cluster multiplicities}
    \label{app:contours}

Equation \ref{eq:racc} presented the contour shape for different cluster multiplicities ($n_{NN}$). Although we only consider $n_{NN} \geq 2$, Eq. \ref{eq:racc} would present the following shape for a single pixel,

 \begin{equation} 
 \mathcal{R}_{\text{acc}}(1, q_\mathrm{min}) \propto \mathcal{I}(q_\mathrm{min}), 
 \end{equation}

 namely, at the single pixel level, the accidental cluster rate reflects the noise rate measured via pedestal events. By considering a cluster of multiplicity $n_{NN}$ = 2:

 \begin{equation} 
 \centering \mathcal{R}_{\text{acc}}(2, q_\mathrm{min}, \delta t_\mathrm{max}) \propto \mathcal{I}^{2}(q_\mathrm{min}) \, \delta t_\mathrm{max}.  
 \end{equation} 
 
\noindent From here, it follows that the contour for $n_{NN}$ = 2 is given by the equation
 
\begin{equation}
    \delta t_\mathrm{max}(q_\mathrm{min}) \propto 
    \frac{\mathcal{R}_{\text{acc}}}{ \mathcal{I}^{2}(q_\mathrm{min})}.
    \label{eq:dt-q}
\end{equation}

\noindent By setting $\mathcal{R}_{\text{acc}}$ to a constant, the acceptable rate of surviving noise is fixed. Similarly, for a cluster of $n_{NN}$ = 3:

\begin{equation}
    \delta t_\mathrm{max}(q_\mathrm{min}) \propto
    \exp\left[ \frac{1}{2}\ln\left( \frac{\mathcal{R}_{\text{acc}}}{ \mathcal{I}^{3}(q_\mathrm{min})} \right) \right],
    \label{eq:dt-q}
\end{equation}

and for a cluster of $n_{NN}$ = 4:

\begin{equation}
    \delta t_\mathrm{max}(q_\mathrm{min}) \propto
    \exp\left[ \frac{1}{3}\ln\left( \frac{\mathcal{R}_{\text{acc}}}{\mathcal{I}^{4}(q_\mathrm{min})} \right) \right].
    \label{eq:dt-q}
\end{equation}

Equation \ref{eq:racc} could be generalised to any number of cluster multiplicity, however, this would greatly increase the complexity of computational implementation. For this reason, in this work we consider a maximum multiplicity of $n_{NN}$ = 4, which has shown to be enough to provide gains in event reconstruction in comparison to the baseline method.

\section{Effects of the accidental cluster rate}
\label{sec:appA}

\begin{figure}[h!]
    \centering
     \includegraphics[width=0.7\hsize]{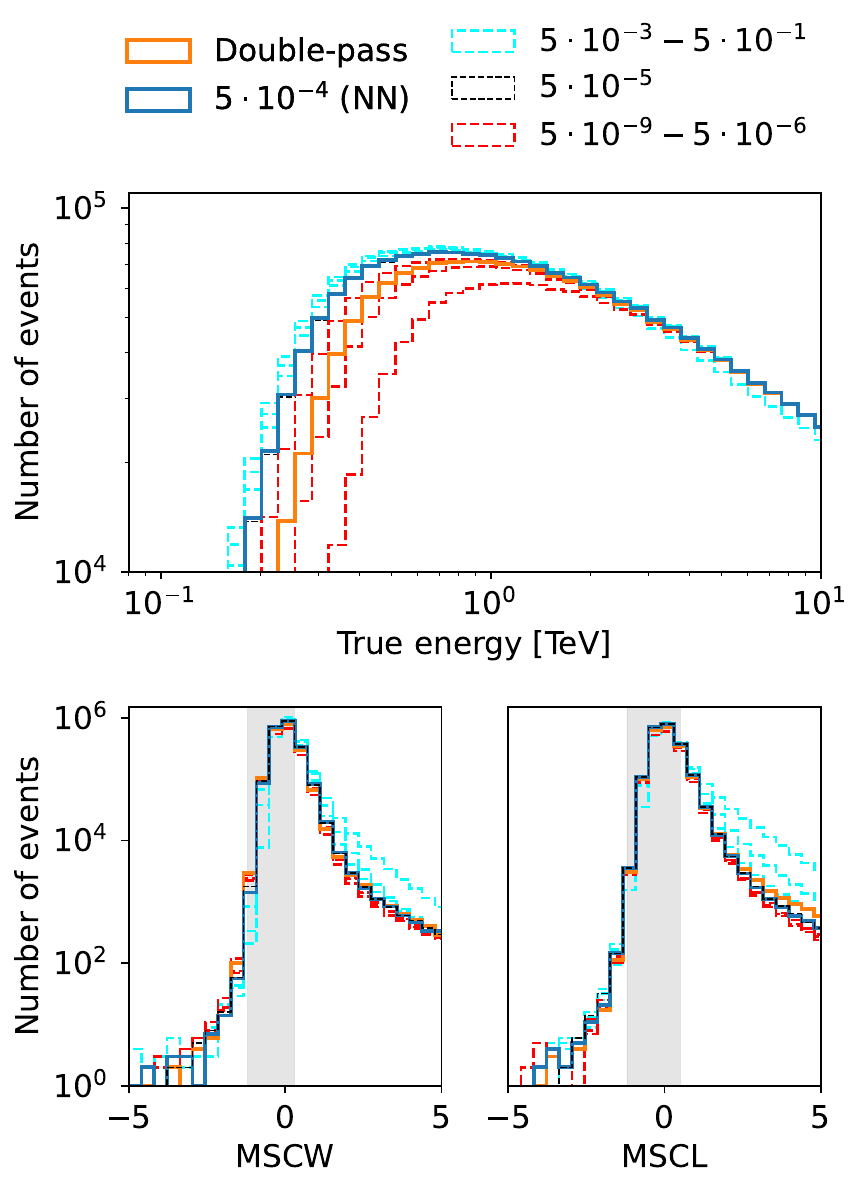}
    \caption{Energy spectrum of simulated gamma-ray events (\textit{top}), distribution of simulated MSCW (\textit{bottom left}) and distribution of simulated MSCL (\textit{bottom right}) for different values of $f_{\text{acc}}$. Simulations were performed at $\phi_z$ = 50 degrees and $\eta$ = 200 MHz. In cyan: $5\cdot 10^{-3} < f_{\mathrm{acc}} < 5\cdot 10^{-1}$. In red: $5\cdot 10^{-9} < f_{\mathrm{acc}} < 5\cdot 10^{-6}$. Blue: histograms for the nominal cleaning. Orange: histograms for the ONN image cleaning, with the chosen $f_{\mathrm{acc}} = 5\cdot 10^{-4}$. The also suitable value of $f_{\mathrm{acc}} = 5\cdot 10^{-5}$ is shown by the black dashed line. For reference, the grey shaded areas represent the supersoft box cuts applied to image parameters at the step of $\gamma$/hadron separation.}
    \label{fig:fprob_50}
\end{figure}

Figure \ref{fig:fprob_50} presents the effect of different accidental cluster rates (and as a consequence, $f_{\rm acc}$) in the distributions of image parameters MSCW and MSCL and on the true energy of simulated showers at large zenith angles (50$^{\circ}$). 

\begin{figure}
   \centering
   \includegraphics[width=0.83\hsize]{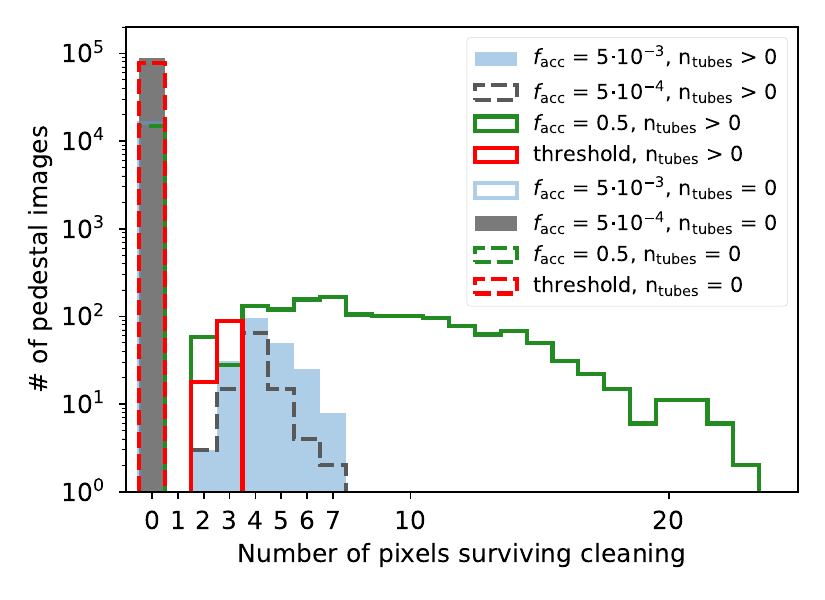}
      \caption{Total number of surviving pixels, $n_\mathrm{tubes}$, for pedestal events cleaned with the ONN method. All four telescopes are considered. The histogram in filled grey (step blue) shows the number of pedestal images with no surviving pixels after the cleaning for $f_{\mathrm{acc}}$ = 5$\cdot$10$^{-4}$ ($f_{\mathrm{acc}}$ = 5$\cdot$10$^{-3}$). The histogram in dashed grey (filled blue) shows the number of pedestal images with at least 2 surviving pixels (minimum multiplicity $n_{NN}$ = 2) after the cleaning for $f_{\mathrm{acc}}$ = 5$\cdot$10$^{-4}$ ($f_{\mathrm{acc}}$ = 5$\cdot$10$^{-3}$).}
         \label{fig:ntubes_peds}
\end{figure}

The number surviving pixels by cleaning pedestal events is shown in Fig. \ref{fig:ntubes_peds}.

\section{Comparison of instrument response functions}
\label{app:irfs}
\begin{figure*}
   \centering
   \resizebox{15cm}{18.58cm}{
   \includegraphics[width=0.97\hsize]{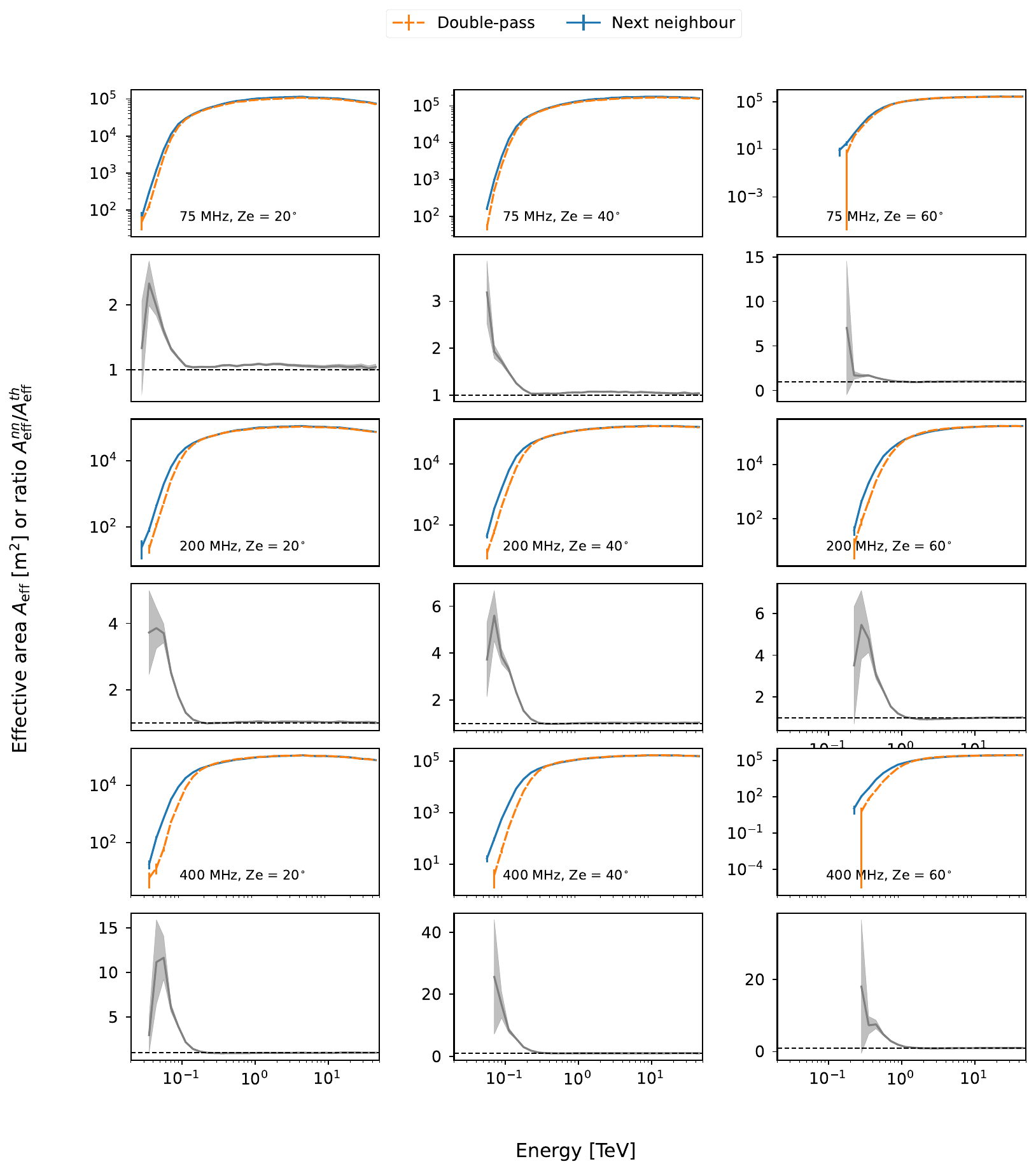}}
      \caption{\textit{From top to bottom}: Effective areas as a function of true shower energy at a noise level of 75 MHz, 200 MHz, and 400 MHz. \textit{From left to right}: Effective areas at a zenith of 20$^{\circ}$, 40$^{\circ}$, and 60$^{\circ}$. Effective areas are given for the ONN image cleaning (in blue) and for the nominal method (in orange). The lower panels represent the ratio of effective areas between the ONN and nominal method, $r_\mathrm{nn/th}$ = $A_{\mathrm{eff}}^{nn}$ / $A_{\mathrm{eff}}^{th}$.}
         \label{fig:effarea}
   \end{figure*}

\begin{figure*}
   \centering
   \resizebox{15cm}{18.58cm}{
   \includegraphics[width=0.97\hsize]{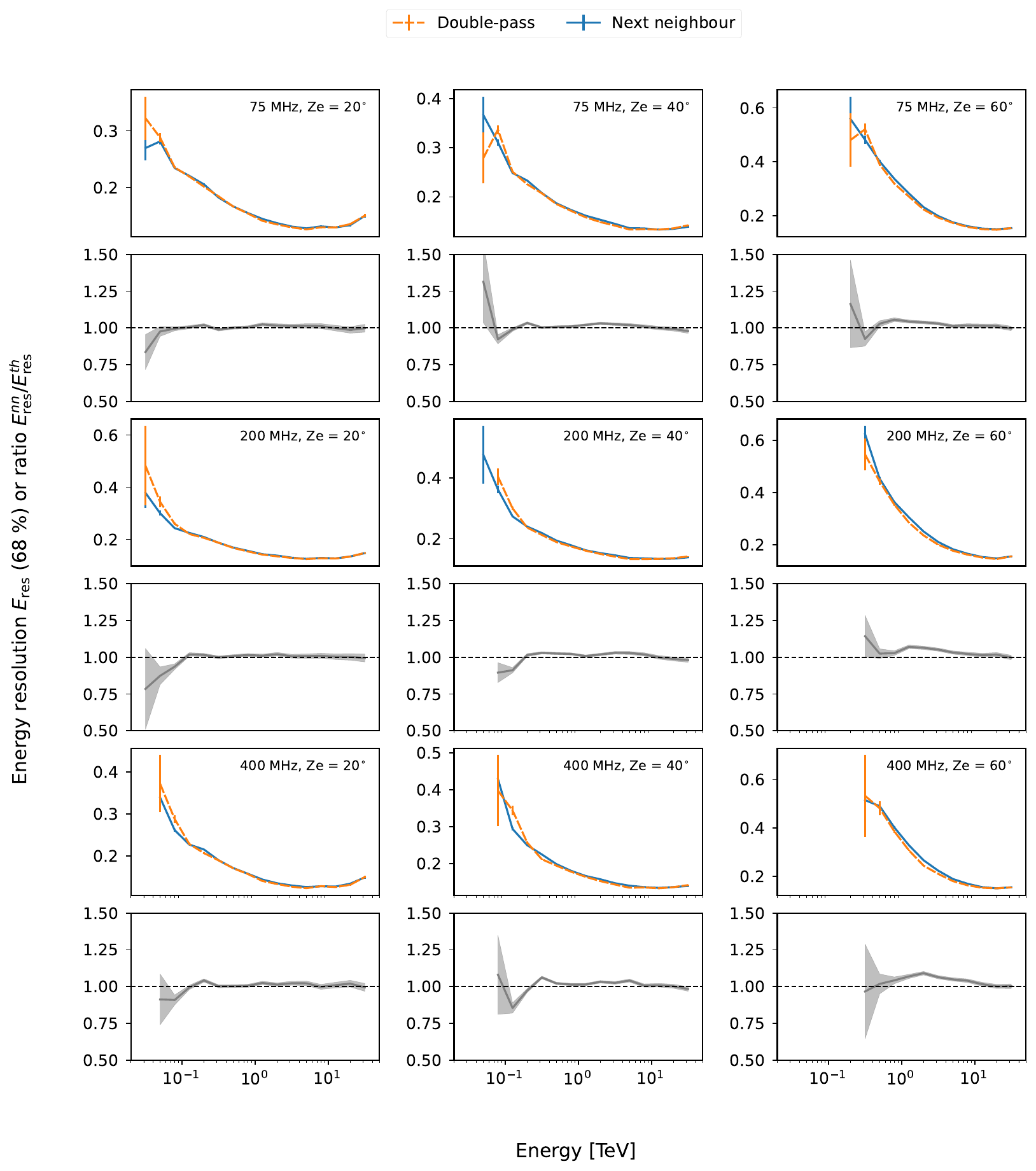}}
      \caption{\textit{From top to bottom}: Energy resolution as a function of true shower energy at a noise level of 75 MHz, 200 MHz, and 400 MHz. \textit{From left to right}: Energy resolution at a zenith of 20$^{\circ}$, 40$^{\circ}$, and 60$^{\circ}$. The energy resolution is shown for the ONN image cleaning (in blue) and for the nominal method (in orange). The lower panels represent the ratio of energy resolution between the ONN and nominal method, $E_{\mathrm{res}}^{nn}$ / $E_{\mathrm{res}}^{th}$.}
         \label{fig:engres}
   \end{figure*}

\begin{figure*}
   \centering
   \resizebox{15cm}{18.58cm}{
   \includegraphics[width=0.97\hsize]{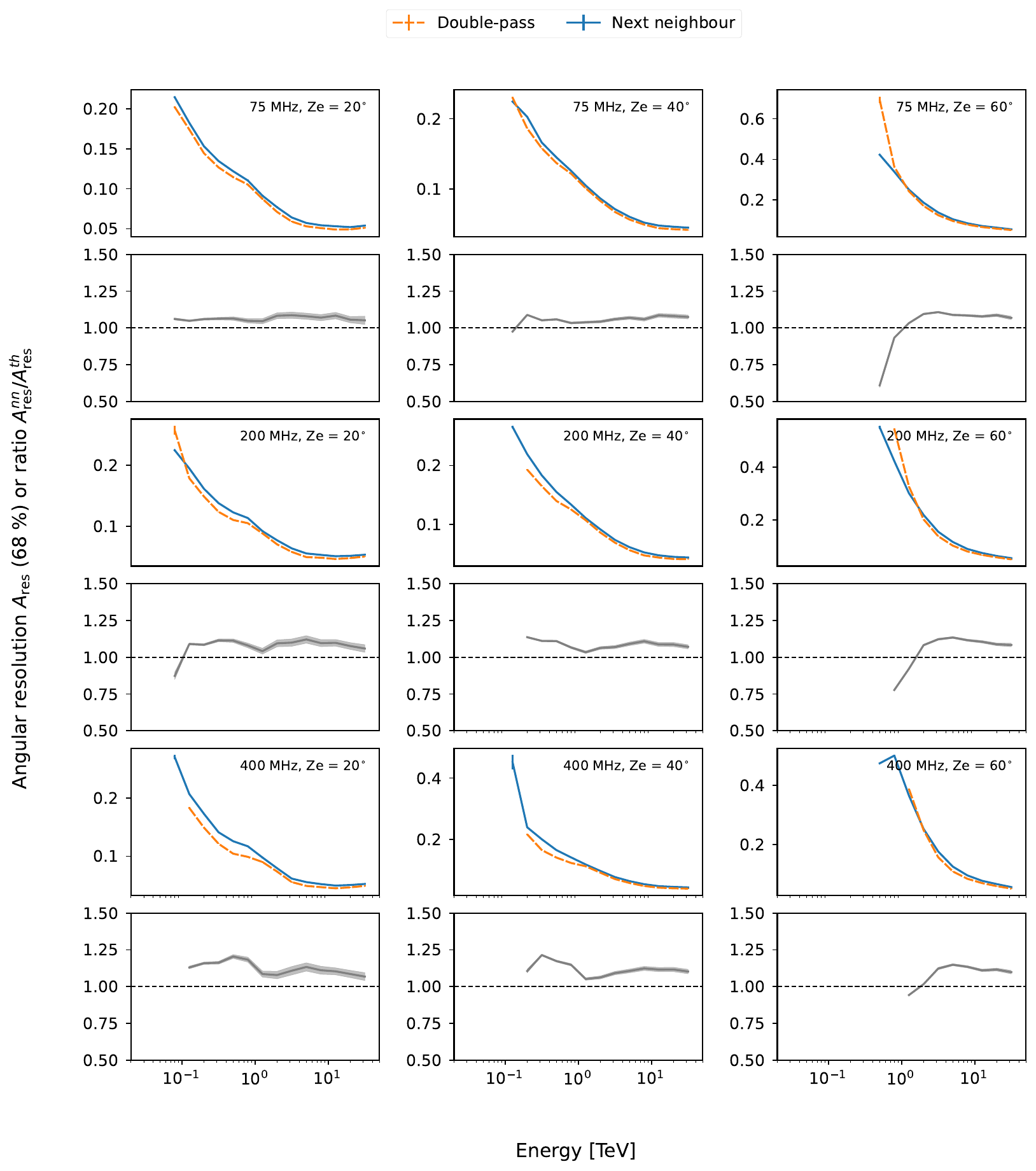}}
      \caption{\textit{From top to bottom}: Angular resolution as a function of true shower energy at a noise level of 75 MHz, 200 MHz, and 400 MHz. \textit{From left to right}: Angular resolution at a zenith of 20$^{\circ}$, 40$^{\circ}$, and 60$^{\circ}$. The angular resolution is shown for the ONN image cleaning (in blue) and for the nominal method (in orange). The lower panels represent the ratio of angular resolution between the ONN and nominal method, $A_{\mathrm{res}}^{nn}$ / $A_{\mathrm{res}}^{th}$.}
         \label{fig:angres}
   \end{figure*}

\begin{figure*}
   \centering
   \resizebox{15cm}{18.58cm}{
   \includegraphics[width=0.97\hsize]{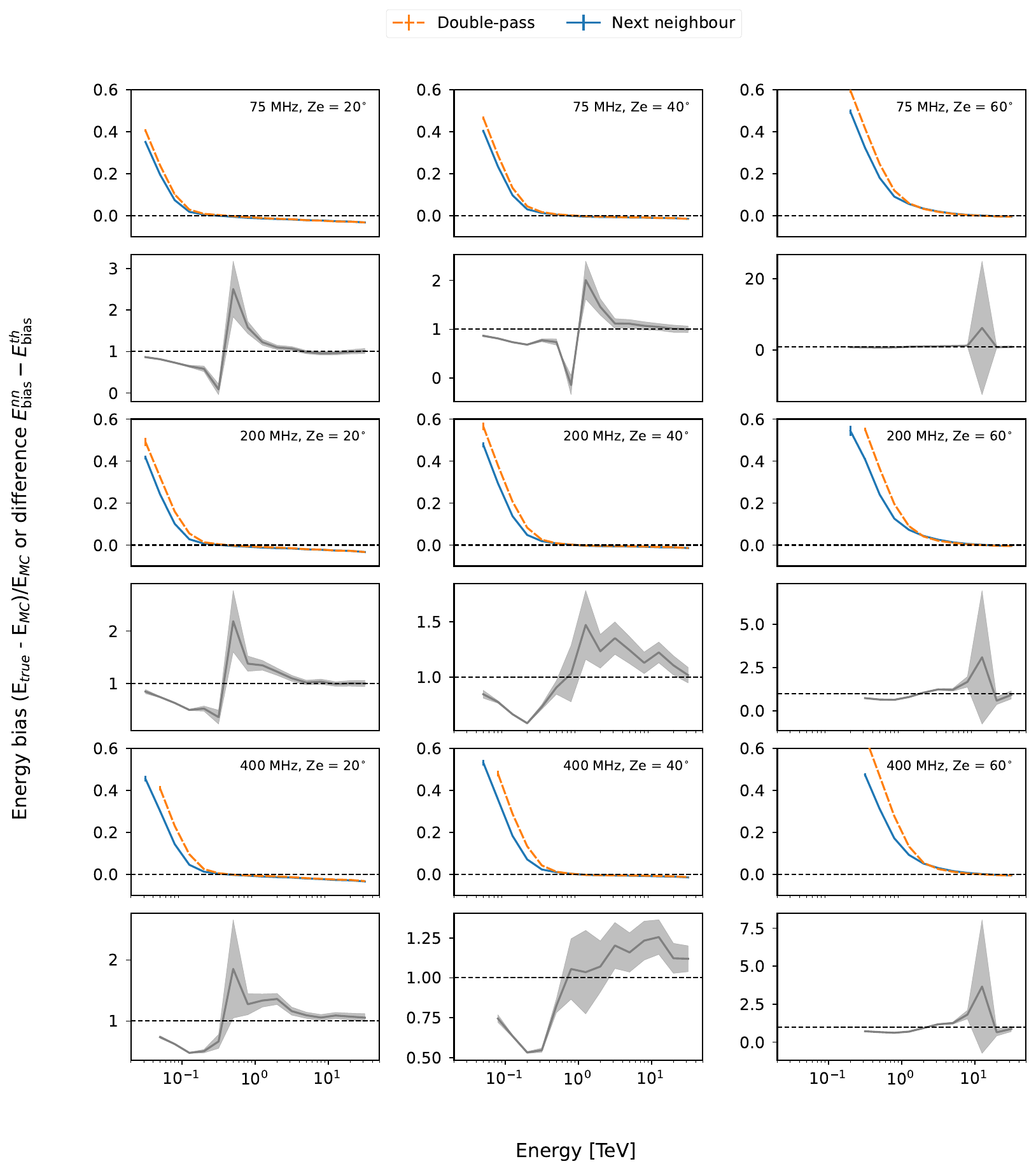}}
      \caption{\textit{From top to bottom}: Energy bias as a function of true shower energy at a noise level of 75 MHz, 200 MHz, and 400 MHz. \textit{\textit{From left to right}}: Energy bias at a zenith of 20$^{\circ}$, 40$^{\circ}$, and 60$^{\circ}$. The energy bias is shown for the ONN image cleaning (in blue) and for the nominal method (in orange). The lower panels represent the ratio of energy bias between the ONN and nominal method, $E_{\mathrm{bias}}^{nn}$ / $E_{\mathrm{bias}}^{th}$.}
         \label{fig:engbias}
   \end{figure*}

Figures \ref{fig:effarea}, \ref{fig:engres}, \ref{fig:angres}, and \ref{fig:engbias} provide, respectively, comparisons between the ONN and baseline cleaning for effective areas, energy resolution, angular resolution and energy bias at different noise levels and zenith angles.

\section{Full Monte Carlo/data comparisons}
\label{app:mc}
\begin{figure*}
    \centering
     \resizebox{18cm}{21cm}
    {\includegraphics[width=2.5\hsize]{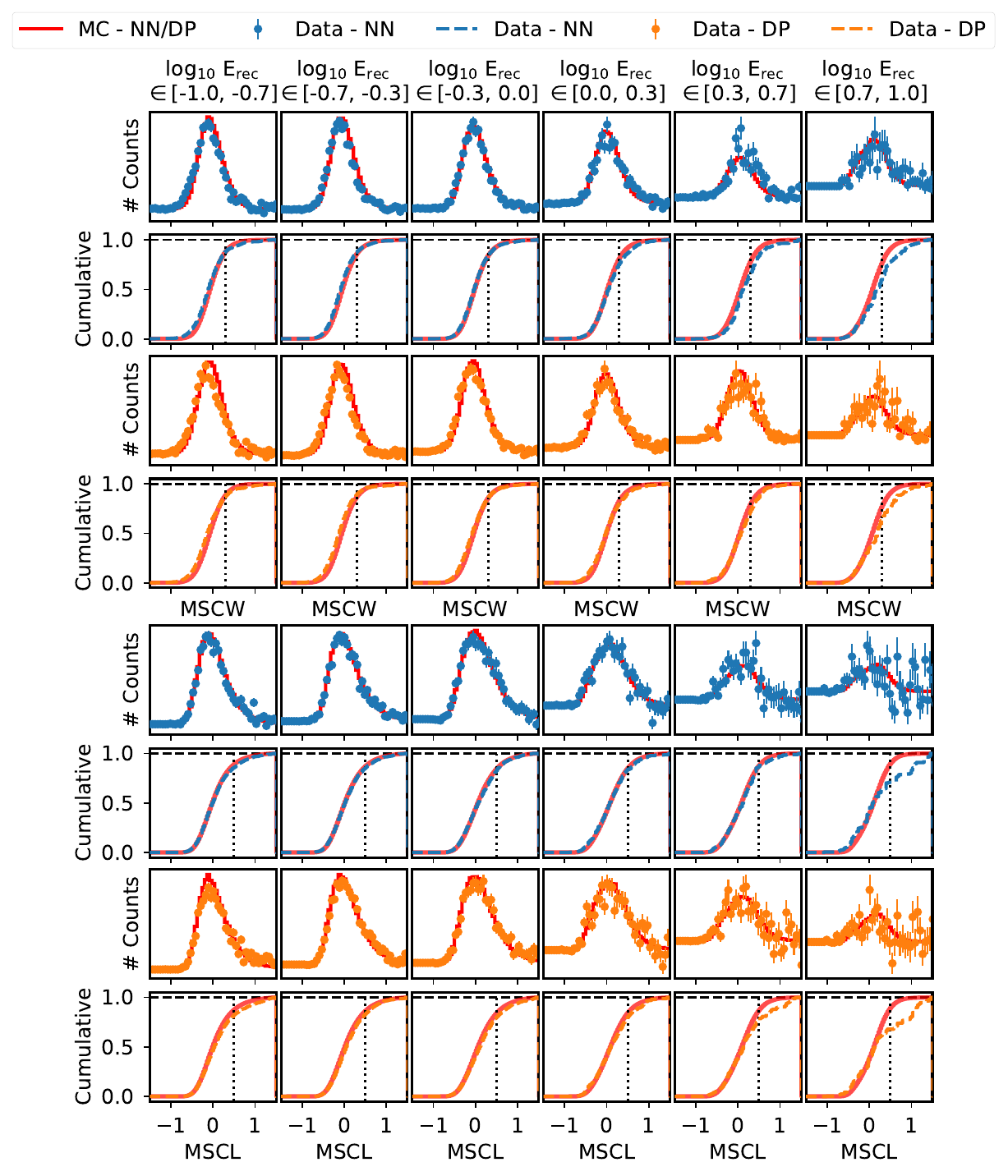}}
    \caption{Spectrum of simulated gamma-ray events (\textit{left}), distribution of simulated MSCW (\textit{centre}) and distribution of simulated MSCL (\textit{right}) for different values of $\mathcal{R}_{\text{acc}}$. Simulations were performed at $\phi_z$ = 20 degrees and $\eta$ = 200 MHz. In cyan: $5\cdot 10^{-3} < f_{\mathrm{acc}} < 5\cdot 10^{0}$. In red: $5\cdot 10^{-8} < f_{\mathrm{acc}} < 5\cdot 10^{-5}$. Blue: histograms for the nominal cleaning. Black: histograms for the ONN image cleaning, with the chosen parameter $f_{\mathrm{acc}} = 5\cdot 10^{-4}$. For reference, the grey shaded regions represent the supersoft box cuts applied to image parameters.}
    \label{fig:full_mc_data_comparison}
\end{figure*}

Figure \ref{fig:full_mc_data_comparison} presents Monte Carlo data comparisons for the distributions of MSCW and MSCL at the energy range $-1.0 \leq {\rm log}_{10} \left( E_{\rm rec}/ {\rm TeV} \right) \leq 1.0$. The corresponding cumulative distribution is provided in energy bins. The vertical dashed lines present the usual cut for $\gamma$/hadron separation.

\end{appendix}
\end{document}